\documentclass{IEEEtran4PSCC}

\usepackage{xcolor}
\usepackage{tikz}
\usetikzlibrary{positioning}
\usetikzlibrary{arrows}
\usetikzlibrary{circuits.ee.IEC}
\usepackage{dblfloatfix}
\usepackage{cite}
\usepackage{amssymb,amsfonts}
\usepackage{amsmath}
\usepackage{booktabs}
\usepackage[font=footnotesize]{caption}
\usepackage[font=footnotesize]{subcaption}
\usepackage[bottom]{footmisc}
\usepackage[english]{babel}
\usepackage{makecell}
\usepgflibrary{shadings}
\usepackage{circledsteps}

\let\proof\relax
\let\endproof\relax
\usepackage{amsthm}

\newtheorem{remark}{Remark}

\usepackage[capitalize,nameinlink,compress]{cleveref}
\crefname{figure}{Fig.}{Figures} % Use Figure instead of Fig.
\crefname{line}{line}{lines} % Make sure line is not capitalized
\crefname{claim}{Claim}{Claims} % Make sure line is not capitalized
\crefname{equation}{}{} % No Eq. for equations
\crefname{problem}{Problem}{Problems}
\crefname{assumption}{Assumption}{Assumptions}

\definecolor{backgroundcolor3}{rgb}{1, 0.493, 0.474}
\definecolor{backgroundcolor}{rgb}{0.122, 0.369, 0.796}
\definecolor{backgroundcolor2}{rgb}{0.4660, 0.6740, 0.1880}
\definecolor{hydro}{rgb}{0.101,0.5,1}
\definecolor{backgroundblue}{rgb}{0, 0.4470, 0.7410}

\makeatletter
\let\old@ps@headings\ps@headings
\let\old@ps@IEEEtitlepagestyle\ps@IEEEtitlepagestyle
\def\psccfooter#1{%
    \def\ps@headings{%
        \old@ps@headings%
        \def\@oddfoot{\strut\hfill#1\hfill\strut}%
        \def\@evenfoot{\strut\hfill#1\hfill\strut}%
    }%
    \def\ps@IEEEtitlepagestyle{%
        \old@ps@IEEEtitlepagestyle%
        \def\@oddfoot{\strut\hfill#1\hfill\strut}%
        \def\@evenfoot{\strut\hfill#1\hfill\strut}%
    }%
    \ps@headings%
}
\makeatother

\usepackage{soul}
\usepackage{framed}
\colorlet{shadecolor}{yellow}
\soulregister\cite7
\soulregister\cref7
\soulregister\ref7
\soulregister\pageref7
\soulregister\eqref7

\renewcommand{\baselinestretch}{0.97}

\begin{document}

\title{\fontdimen2\font=0.69ex Dynamic Ancillary Services: From Grid Codes to Transfer Function-Based Converter Control\vspace{-2mm}}

\author{\IEEEauthorblockN{Verena Häberle\IEEEauthorrefmark{1},
 Linbin Huang\IEEEauthorrefmark{1},
 Xiuqiang He\IEEEauthorrefmark{1}, 
 Eduardo Prieto-Araujo\IEEEauthorrefmark{2} and
 Florian Dörfler\IEEEauthorrefmark{1}}
 \IEEEauthorblockA{\IEEEauthorrefmark{1} Automatic Control Laboratory, ETH Zurich, 8092 Zurich, Switzerland}
 \IEEEauthorblockA{\IEEEauthorrefmark{2} CITCEA, Universitat Politècnica de Catalunya, 08028 Barcelona, Spain}
$\vspace{-1mm}$
 }

\maketitle

\begin{abstract}\fontdimen2\font=0.6ex
Conventional grid-code specifications for dynamic ancillary services provision such as fast frequency and voltage regulation are typically defined by means of piece-wise linear step-response capability curves in the time domain. However, although the specification of such time-domain curves is straightforward, their practical implementation in a converter-based generation system is not immediate, and no customary methods have been developed yet. In this paper, we thus propose a systematic approach for the practical implementation of piece-wise linear time-domain curves to provide dynamic ancillary services by converter-based generation systems, while ensuring grid-code and device-level requirements to be reliably satisfied. Namely, we translate the piece-wise linear time-domain curves for active and reactive power provision in response to a frequency and voltage step change into a desired rational parametric transfer function in the frequency domain, which defines a dynamic response behavior to be realized by the converter. The obtained transfer function can be easily implemented e.g. via a proportional-integral (PI)-based matching control in the power loop of standard converter control architectures. We demonstrate the performance of our method in numerical grid-code compliance tests, and reveal its superiority over classical droop and virtual inertia schemes which may not satisfy the grid codes due to their structural limitations.
\end{abstract}

\thanksto{\noindent This work was supported by the European Union's Horizon 2020 and 2023 research and innovation programs (Grant Agreement Numbers 883985 and 101096197). Corresponding author's email address: verenhae@ethz.ch.}

\section{Introduction}\fontdimen2\font=0.6ex
%structure of introduction
%Paragraph 1: Motivation. At a high level, what is the problem area you are working in and why is it important?
Today's grid-code specifications for dynamic ancillary services provision such as fast frequency and voltage control are typically defined by means of a prescribed time-domain step-response characteristic \cite{european2016commission,oyj2021technical,Eirgrid2018}. As an example, the European network code \cite{european2016commission}, which is adopted in most European national grid codes, specifies the active power provision for frequency containment reserve (FCR) in response to a frequency step change by a piece-wise linear time-domain capability curve, where the required active power response should be satisfied at or above the curve. Likewise, the dynamic response of reactive power for voltage control is defined via time specifications in response to a voltage step change. Recently, also modern grid codes (e.g., Finland \cite{oyj2021technical}, Ireland \cite{Eirgrid2018}) define the activation of fast frequency reserves (FFR) or synthetic inertia via piece-wise linear active power capability curves in the time domain. 

%Paragraph 2: What is the specific problem considered in this paper? This paragraph narrows down the topic area of the paper.
Although the specification of the piece-wise linear time-domain curves in today's grid codes is straightforward, their practical implementation in a converter-based generation system is not immediate, and there is a lack of systematic approaches. In particular, the classical droop and virtual inertia control \cite{poolla2019placement,ferreira2021control,stanojev2022improving} may not be able to achieve the required specifications due to their fixed controller structure (therefore may not pass the grid-code compliance tests), or would require a cumbersome trial-and-error tuning procedure by adding filters to approximately satisfy the requirements and the local device-level constraints. In this regard, today's industrial practice to implement time-domain grid-code curves is usually very ad-hoc, e.g., often relying on methods similar to open-loop trajectory commands \cite{ruttledge2015emulated}, varying-gain \cite{wu2018state,hwang2016dynamic}, or look-up table schemes\cite{clark2010modeling}. In particular, the work in \cite{ruttledge2015emulated} discusses a fixed trajectory response approach where for any deviation in frequency beyond a certain threshold, a fixed response shape is triggered. Such a fixed response shape could then be used to replicate a desired time-domain grid-code curve. On the other hand, the work in \cite{wu2018state,hwang2016dynamic} proposes variable gain methods (e.g., based on variable droop gains) to adjust and optimize the dynamic response of wind turbines for frequency regulation. Also with such an approach, one could approximate the desired time-domain grid-code response behavior by exploiting the flexibility of the varying-gain control. Ultimately, the work in \cite{ruttledge2015emulated} discusses some look-up table scheme for open-loop reactive power control, which could again be used to reproduce the dynamic response behavior of piece-wise linear time-domain grid-code curves. Nevertheless, for all the methods outlined in \cite{ruttledge2015emulated,wu2018state,hwang2016dynamic,clark2010modeling}, the design, implementation, and tuning are in general quite time-consuming and rigid, which impedes its practical applicability.
 
%Paragraph 3: "In this paper, we show that ...". This is the key paragraph in the intro - you summarize, in one paragraph, what are the main contributions of your paper. 
In this work, we propose a systematic way to provide dynamic ancillary services with converter-based generation systems, which ensures that desired grid-code and device-level requirements are reliably satisfied. In particular, we translate the piece-wise linear time-domain grid-code curves for active and reactive power provision into a \textit{desired} rational transfer function matrix in the frequency domain, which defines a tractable response behavior to be realized by the converter. Since the conventional cascaded control structure of power converters is typically designed for tracking active and reactive power reference signals, it is well-suited to further include a reference model as given by the desired transfer function, and thus enables a simple matching control implementation. 

Finally, the proposed method is versatile to match any piece-wise linear time-domain capability curve, and thus also allows for more complex grid-code specifications in the future, e.g., to address grid-forming requirements. Beyond that, our results might even inspire a change of perspective in such a way that there will be an immediate transfer function-based formulation of future grid codes specified in the frequency domain.

\section{Preliminaries: Grid-Code Specifications}\label{sec:grid_code_curves}\fontdimen2\font=0.6ex
We consider conventional grid-code specifications for dynamic frequency and voltage regulation in grid-following converters (i.e., reserve units), which are typically defined by means of a prescribed time-domain step-response characteristic  \cite{european2016commission,oyj2021technical,Eirgrid2018}. Different schematic examples of such grid-code specifications are depicted in \cref{fig:gridcode_examples} and presented in the following. 
t\subsubsection*{Example 1 --- FCR Provision} The first example in \cref{fig:fcr_gridcode} is extracted from the European network code \cite{european2016commission}, which is adopted in most European national grid codes, and the plotted piece-wise linear time-domain capability curve is used to specify the active power provision for frequency containment reserve (FCR) in response to a frequency step change. In particular, the FCR-providing reserve unit has to deliver a certain \textit{normalized} active power FCR capacity $|\Delta p_\mathrm{fcr}^\mathrm{n}|$ in accordance with an initial delay time $t_\mathrm{i}^\mathrm{fcr}$ and a full activation time $t_\mathrm{a}^\mathrm{fcr}$, where the normalized FCR capacity $|\Delta p_\mathrm{fcr}^\mathrm{n}|$ is conventionally given by the allocated active power droop gain $D_\mathrm{p}>0$ for a \textit{unit} frequency step change\footnote{To introduce the formalism of the grid-code specifications and translate them into rational parametric transfer functions in \cref{sec:grid_code_curves,sec:grid_code_2_tf}, we consider \textit{normalized} active and reactive power capability curves in response to (practically unrealistic) \textit{unit} step changes in frequency and voltage, i.e., $|\Delta f| = 1$ p.u. and $|\Delta v| = 1$ p.u., respectively. In a practical implementation setup, however, the capability curves are scaled with a non-unity frequency and voltage step input in order to obtain reasonable power capacities which are realizable by a converter-based generation system (cf. \cref{sec:tf_based_control,sec:case_studies}).} $|\Delta f|=1$ p.u. as
$|\Delta p_\mathrm{fcr}^\mathrm{n}| := \tfrac{1}{D_\mathrm{p}}|\Delta f|=\tfrac{1}{D_\mathrm{p}}$. The time parameters $t_\mathrm{i}^\mathrm{fcr}$ and $t_\mathrm{a}^\mathrm{fcr}$, in turn, allow for some flexibility of the reserve unit, as they are not fixed to a particular value, but only have to satisfy the following \color{backgroundcolor2}grid-code \color{black} and \color{magenta}device-level \color{black} requirements \cite{european2016commission}, i.e., 
\begin{subequations}\label{eq:grid_code_req_fcr}
\begin{align}\label{eq:grid_code_req_fcr1}
    0\leq t_\mathrm{i}^\mathrm{fcr} &\leq \color{backgroundcolor2}T_\mathrm{i,max}^\mathrm{fcr}\color{black}\\\label{eq:grid_code_req_fcr2}
    t_\mathrm{i}^\mathrm{fcr} \leq t_\mathrm{a}^\mathrm{fcr} &\leq \color{backgroundcolor2}T_\mathrm{a,max}^\mathrm{fcr}\color{black}\\\label{eq:grid_code_req_fcr3}
    |\Delta p_\mathrm{fcr}^\mathrm{n}|&\leq \left(t_\mathrm{a}^\mathrm{fcr}-t_\mathrm{i}^\mathrm{fcr}\right)\cdot \color{magenta}R_\mathrm{max}^\mathrm{p}\color{black},
\end{align}
\end{subequations}
where \color{backgroundcolor2}$T_\mathrm{i,max}^\mathrm{fcr}$ \color{black} and \color{backgroundcolor2}$T_\mathrm{a,max}^\mathrm{fcr}$ \color{black} are the maximum admissible FCR initial delay and activation times, and \color{magenta} $R_\mathrm{max}^\mathrm{p}$ \color{black} is the normalized maximal active power ramping rate of the reserve unit. 

\subsubsection*{Example 2 --- Voltage Control} 
\begin{figure}[t!]
\centering
    \begin{subfigure}{0.23\textwidth}
        \vspace{-1mm}
        \centering
        \resizebox{1.05\textwidth}{!}{
\begin{tikzpicture}
\draw[-latex] (-2.8,0.9) -- (-2.8,3.1); 
\draw [-latex](-2.9,1) -- (1.1,1);
\node [scale=0.9]at (-3.2,3) {$|\Delta p|$};
\node [scale=0.9]at (1.1,0.8) {$t$};
\draw [ultra thick,black!40](-2.8,1)--(-2.4,1) -- (-0.1,2) -- (0.9,2);
\draw[dashed] (-0.1,2) -- (-0.1,1);
\node [scale=0.9]at (-0.1,0.75) {$t_\mathrm{a}^\mathrm{fcr}$};

\draw[dotted,black!60](-0.1,2)  -- (-2.8,2);
\node [scale=0.9,black!60] at (-3.35,2) {$|\Delta p_\mathrm{fcr}^\mathrm{n}|$};
%\node [scale=0.7,black!60] at (-3.3,1.88) {capacity};

\node [scale=0.9] at (-2.4,0.75) {$t_\mathrm{i}^\mathrm{fcr}$};
\end{tikzpicture}
}
        \vspace{-9.5mm}
        \caption{Normalized active power capability curve for FCR provision after a unit frequency step change \cite{european2016commission}.}
        \vspace{2mm}
        \label{fig:fcr_gridcode}
    \end{subfigure}
    \hspace{2mm}
       \begin{subfigure}{0.23\textwidth}
          \centering
          \vspace{-1mm}
    \resizebox{1.05\textwidth}{!}{
\begin{tikzpicture}
\draw[-latex] (2.3,0.9) -- (2.3,3.1); 
\draw [-latex](2.2,1) -- (6.2,1);
\node [scale=0.9]at (1.9,3) {$|\Delta q|$};
\node [scale=0.9]at (6.2,0.8) {$t$};

\draw[dashed] (2.9,2.1) node (v2) {} -- (2.9,1);

\draw [dashed](4.2,2.4) -- (4.2,1); 
\draw [dotted,black!60](4.2,2.4) -- (2.3,2.4);
\node  [scale=0.9,black!60] at (1.72,2.45) {$|\Delta q_{100}^\mathrm{n}|$};

\node [scale=0.9] at (2.9,0.75) {$t_\mathrm{90}^\mathrm{vq}$};

\node [scale=0.9]at (4.2,0.75) {$t_\mathrm{100}^\mathrm{vq}$};

\draw[dotted,black!60](2.9,2.1)  -- (2.3,2.1);

\draw [ultra thick, black!40](2.3,1) --  (2.9,2.1) -- (4.2,2.4) -- (6,2.4) ;

\node   [scale=0.9,black!60] at (1.78,2.05) {$|\Delta q_{90}^\mathrm{n}|$};
\end{tikzpicture}
}
             \vspace{-9.5mm}
    \caption{Normalized reactive power capability curve for voltage control after a unit voltage step change \cite{european2016commission}.}
        \vspace{2mm}
    \label{fig:vq_gridcode}
    \end{subfigure}
        \begin{subfigure}{0.23\textwidth}
        \centering
        \resizebox{1.05\textwidth}{!}{
\begin{tikzpicture}
\draw[-latex] (2.3,0.9) -- (2.3,3.6); 
\draw [-latex](2.2,1) -- (6.2,1);
\node [scale=0.9]at (1.9,3.5) {$|\Delta p|$};
\node [scale=0.9]at (6.2,0.8) {$t$};
\draw [ultra thick,black!40](2.3,1) -- (2.6,3) --(3.9,2.3)-- (4.6,1)--(6,1);
\draw[dashed] (2.5,2.3) node (v2) {} -- (2.5,1);
\node [scale=0.9]at (3.9,0.75) {$t_\mathrm{d}^\mathrm{ffr}$};

\draw [dashed](3.9,2.3) -- (3.9,1); 
\draw [dotted,black!60](3.9,2.3) -- (2.25,2.3);
\node  [scale=0.9,black!60] at (1.775,2.35) {$|\Delta p_\mathrm{ffr}^\mathrm{n}|$};

\node [scale=0.9] at (2.5,0.75) {$t_\mathrm{a}^\mathrm{ffr}$};

\draw [dotted,black!60](2.6,3) -- (2.3,3);

\node [scale=0.9]at (4.6,0.75) {$t_\mathrm{r}^\mathrm{ffr}$};

\node  [scale=0.9,black!60] at (1.55,3) {$|\Delta p_\mathrm{ffr}^\mathrm{peak,n}|$};
%\node [scale=0.7,black!60] at (1.8,2.18) {capacity};
\end{tikzpicture}
}
         \vspace{-9.5mm}
        \caption{Normalized active power capability curve for FFR provision after a unit frequency step change \cite{oyj2021technical}.}
        \label{fig:ffr_gridcode}
    \end{subfigure}
        \hspace{2mm}
    \begin{subfigure}{0.23\textwidth}
          \centering
        \resizebox{1.05\textwidth}{!}{
\begin{tikzpicture}
\draw[-latex] (2.3,0.9) -- (2.3,3.8); 
\draw [-latex](2.2,1) -- (6.2,1);
\node [scale=0.9] at (1.9,3.5) {$|\Delta p|$};
\node [scale=0.9]at (6.2,0.8) {$t$};
\draw [black!40](2.3,1) -- (2.6,3.1) --(3.9,2.3)-- (4.6,1)--(6,1);
\draw[dashed] (2.5,2.3) node (v2) {} -- (2.5,1);
\node [scale=0.9]at (3.9,0.75) {$t_\mathrm{d}^\mathrm{ffr}$};

\draw [dashed](3.9,2.3) -- (3.9,1); 
\draw [dotted,black!60](3.9,2.3) -- (2.22,2.3);
%\node  [scale=0.7,black!60] at (1.75,2.3) {FFR cap.};
\node [scale=0.9,black!60] at (1.77,2.35) {$|\Delta p_\mathrm{ffr}^\mathrm{n}|$};

\node [scale=0.9] at (2.4,0.75) {$t_\mathrm{a}^\mathrm{ffr}$};
\draw[dashed] (5,2) -- (5,1);
\node [scale=0.9]at (5,0.75) {$t_\mathrm{a}^\mathrm{fcr}$};

\draw[dotted,black!60](5,2)  -- (2.22,2);
\draw [black!40](2.3,1) node (v1) {}--(2.7,1) -- (5,2) node (v3) {} -- (6,2) node (v4) {};
\node [scale=0.9]at (4.6,0.75) {$t_\mathrm{r}^\mathrm{ffr}$};
\draw [ultra thick, black!40](2.3,1) --  (2.6,3.1) -- (2.7,3.05) -- (3.9,2.85) -- (4.6,1.82) -- (5,2) -- (6,2) ;
\node [scale=0.9] at (2.8,0.75) {$t_\mathrm{i}^\mathrm{fcr}$};
%\node   [scale=0.7,black!60]at (1.75,2) {FCR cap.};
\node [scale=0.9,black!60] at (1.75,1.95) {$|\Delta p_\mathrm{fcr}^\mathrm{n}|$};

\node [scale=0.9,black!60]at (1.53,3.1) {$|\Delta p_\mathrm{ffr}^\mathrm{peak,n}|$};
\draw[dotted,black!60] (2.6,3.1) -- (2.27,3.1);
\end{tikzpicture}
}
         \vspace{-9.5mm}
        \caption{Normalized superimposed capability curve for FFR-FCR provision after a unit frequency step change.}
        \label{fig:ffr_fcr_superimposed_gridcode}
    \end{subfigure}
    \caption{Examples of piece-wise linear time-domain capability curves to provide dynamic ancillary services in different grid codes (simplified).}
     \vspace{-4mm}
    \label{fig:gridcode_examples}
\end{figure}
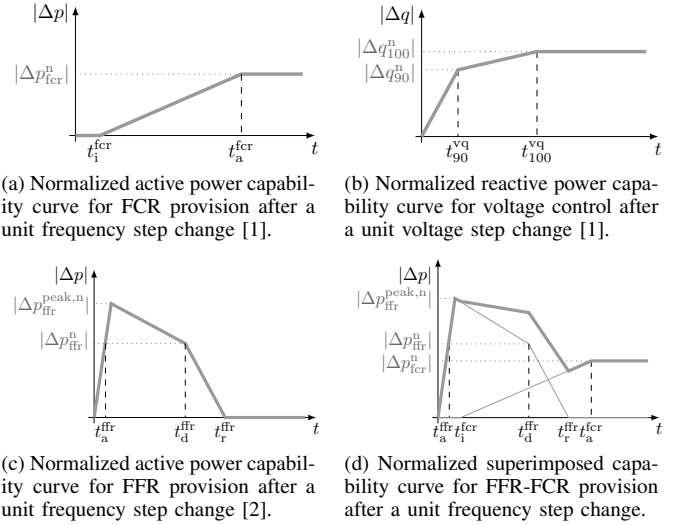

Likewise, as illustrated in \cref{fig:vq_gridcode}, the European network code defines the dynamic activation of a certain reactive power capacity in response to a voltage step change, where the \textit{normalized} reactive power capacity levels $|\Delta q_{90}^\mathrm{n}|$ of 90\% and $|\Delta q_{100}^\mathrm{n}|$ of 100\% have to be achieved in accordance with the times $t_{90}^\mathrm{vq}$ and $t_{100}^\mathrm{vq}$, respectively. The normalized reactive power capacity levels are conventionally given by the allocated reactive power droop gain $D_\mathrm{q}>0$ during a \textit{unit} voltage step change $|\Delta v|=1$ p.u. as $|\Delta q_\mathrm{100}^\mathrm{n}| := \tfrac{1}{D_\mathrm{q}}|\Delta v|=\tfrac{1}{D_\mathrm{q}}$ and $|\Delta q_\mathrm{90}^\mathrm{n}|=0.9\cdot |\Delta q_\mathrm{100}^\mathrm{n}|$, while the time parameters $t_{90}^\mathrm{vq}$ and $t_{100}^\mathrm{vq}$ have to satisfy the following \color{backgroundcolor2} grid-code \color{black} and \color{magenta} device-level \color{black} requirements \cite{european2016commission}, i.e.,
\begin{subequations}\label{eq:grid_code_req_reactive_power}
    \begin{align}\label{eq:grid_code_req_reactive_power1}
    0\leq t_{90}^\mathrm{vq} &\leq \color{backgroundcolor2} T_\mathrm{90,max}^\mathrm{vq}\color{black}\\\label{eq:grid_code_req_reactive_power2}
    t_{90}^\mathrm{vq} \leq t_{100}^\mathrm{vq} &\leq \color{backgroundcolor2}T_\mathrm{100,max}^\mathrm{vq}\color{black}\\\label{eq:grid_code_req_reactive_power3}
    |\Delta q_{90}^\mathrm{n}| &\leq t_{90}^\mathrm{vq} \cdot \color{magenta}R_\mathrm{max}^\mathrm{q}\color{black}\\\label{eq:grid_code_req_reactive_power4}
    0.1\cdot |\Delta q_{100}^\mathrm{n}| &\leq \left(t_{100}^\mathrm{vq}-t_{90}^\mathrm{vq}\right) \cdot \color{magenta} R_\mathrm{max}^\mathrm{q}
    \color{black}
    \end{align}
\end{subequations}
where \color{backgroundcolor2} $T_\mathrm{90,max}^\mathrm{vq}$ \color{black} and \color{backgroundcolor2} $T_\mathrm{100,max}^\mathrm{vq}$ \color{black} are the maximum admissible 90\% and 100\% activation times for the reactive power capacity provision, respectively, and \color{magenta} $R_\mathrm{max}^\mathrm{q}$ \color{black} is the normalized maximum reactive power ramping rate of the reserve unit. 

\subsubsection*{Example 3 --- FFR Provision} To ensure frequency regulation on faster time scales to counteract the effect of reduced inertial response in future power systems, modern grid codes (e.g., Finland \cite{oyj2021technical}, Ireland \cite{Eirgrid2018}) recently also specify the activation of fast frequency reserves (FFR) or synthetic inertia via piece-wise linear active power capability curves in the time domain. An example of such a \textit{normalized} active power capability curve for FFR provision after a \textit{unit} step change in frequency is shown in \cref{fig:ffr_gridcode}. More specifically, the FFR-providing reserve unit has to deliver a certain normalized active power FFR capacity $|\Delta p_\mathrm{ffr}^\mathrm{n}|$ after an activation time $t_\mathrm{a}^\mathrm{ffr}$, which has to remain activated for a particular support duration $t_\mathrm{d}^\mathrm{ffr}-t_\mathrm{a}^\mathrm{ffr}$. The reserve unit can return to recovery at time $t_\mathrm{r}^\mathrm{ffr}$ after the support duration has elapsed. The FFR capacity $|\Delta p_\mathrm{ffr}^\mathrm{n}|$ is usually specified during prequalification tests of the reserve unit \cite{oyj2021technical}, where a unit step change in frequency $|\Delta f|=1$ p.u. is correlated with a particular amount of active power that can be provided in the reserve market. Similar to droop gains, we define a constant correlation factor $K_\mathrm{p}>0$ to calculate the FFR capacity as $|\Delta p_\mathrm{ffr}^\mathrm{n}| := \tfrac{1}{K_\mathrm{p}} |\Delta f|=\tfrac{1}{K_\mathrm{p}}$. Moreover, the active power change of the reserve unit may exceed the FFR capacity $|\Delta p_\mathrm{ffr}^\mathrm{n}|$ up to a certain overdelivery $|\Delta p_\mathrm{ffr}^\mathrm{peak,n}|$. In practice, the above time parameters $t_\mathrm{a}^\mathrm{ffr}$, $t_\mathrm{d}^\mathrm{ffr}$, $t_\mathrm{r}^\mathrm{ffr}$ and the overdelivery $|\Delta p_\mathrm{ffr}^\mathrm{peak,n}|$ are subject to constraints from \color{backgroundcolor2} grid-code \color{black} and \color{magenta} device-level \color{black} requirements as \cite{oyj2021technical,Eirgrid2018}
\begin{subequations}\label{eq:grid_code_req_ffr}
\begin{align}\label{eq:grid_code_req_ffr1}
0 \leq t_\mathrm{a}^\mathrm{ffr} &\leq \color{backgroundcolor2}T_\mathrm{a,max}^\mathrm{ffr}\color{black}\\\label{eq:grid_code_req_ffr2}
|\Delta p_\mathrm{ffr}^\mathrm{n}| & \leq t_\mathrm{a}^\mathrm{ffr}\cdot \color{magenta} R_\mathrm{max}^\mathrm{p} \color{black}\\\label{eq:grid_code_req_ffr3}
        \color{magenta} T_\mathrm{d,max}^\mathrm{ffr} \color{black} \geq t_\mathrm{d}^\mathrm{ffr}-t_\mathrm{a}^\mathrm{ffr} &\geq \color{backgroundcolor2}T_\mathrm{d,min}^\mathrm{ffr}\color{black}\\\label{eq:grid_code_req_ffr4}
           \color{magenta} T_\mathrm{r,max}^\mathrm{ffr} \color{black} \geq t_\mathrm{r}^\mathrm{ffr}-t_\mathrm{d}^\mathrm{ffr} &\geq \color{backgroundcolor2} T_\mathrm{r,min}^\mathrm{ffr} \color{black}\\\label{eq:grid_code_req_ffr5}
           \color{backgroundcolor2} |\Delta p_\mathrm{ffr}^\mathrm{n}| \color{black} \leq|\Delta p_\mathrm{ffr}^\mathrm{peak,n}| &\leq \text{min}\left \{  \color{magenta}M_\mathrm{max}^\mathrm{p} \color{black} , \color{backgroundcolor2} x_\mathrm{peak}^\mathrm{ffr}\cdot |\Delta p_\mathrm{ffr}^\mathrm{n}| \color{black} \right\},
\end{align}
\end{subequations}
where \color{backgroundcolor2} $T_\mathrm{a,max}^\mathrm{ffr}$ \color{black} is the maximum admissible full activation time for the FFR provision. The FFR support duration $t_\mathrm{d}^\mathrm{ffr}-t_\mathrm{a}^\mathrm{ffr}$ is lower and upper bounded by the minimum and maximum support duration \color{backgroundcolor2} $T_\mathrm{d,min}^\mathrm{ffr}\gg T_\mathrm{a,max}^\mathrm{ffr}$ \color{black} and \color{magenta} $T_\mathrm{d,max}^\mathrm{ffr}$\color{black}, respectively. Likewise, the minimum and maximum return-to-recovery times are given by \color{backgroundcolor2} $T_\mathrm{r,min}^\mathrm{ffr}\geq T_\mathrm{a,max}^\mathrm{ffr}$ \color{black} and \color{magenta} $T_\mathrm{r,max}^\mathrm{ffr}$\color{black}, respectively. The FFR overdelivery $|\Delta p_\mathrm{ffr}^\mathrm{peak,n}|$ must not exceed the reserve unit's normalized maximum active power peak capacity \color{magenta} $M_\mathrm{max}^\mathrm{p}$, \color{black} as well as the maximum tolerable overdelivery \color{backgroundcolor2} $x_\mathrm{peak}^\mathrm{ffr}\cdot|\Delta p_\mathrm{ffr}^\mathrm{n}|$\color{black}, where the exceedance factor $\color{backgroundcolor2} x_\mathrm{peak}^\mathrm{ffr}\color{black}\in[1,2]$ is specified in the grid code. Finally, \color{magenta} $R_\mathrm{max}^\mathrm{p}$ \color{black} is again the reserve unit's normalized maximum active power ramping rate.
\renewcommand{\arraystretch}{1.2}
\begin{table}[t!]\scriptsize
    \centering
     \setlength{\tabcolsep}{1mm}
           \caption{Exemplary values for \color{backgroundcolor2} grid-code \color{black} specifications \cite{european2016commission,oyj2021technical,Eirgrid2018}.}
               \begin{tabular}{c||c|c}
     \toprule
          Parameter & Symbol & Value  \\ \hline
           Maximum admissible initial delay time for FCR provision & \color{backgroundcolor2}$T_\mathrm{i,max}^\mathrm{fcr}$ & 2 s \\
           Maximum admissible full activation time for FCR provision & \color{backgroundcolor2}$T_\mathrm{a,max}^\mathrm{fcr}$& 30 s\\ \hline
           Maximum admissible 90\% reactive power activation time & \color{backgroundcolor2}$T_\mathrm{90,max}^\mathrm{vq}$& 5 s\\
           Maximum admissible 100\% reactive power activation time & \color{backgroundcolor2}$T_\mathrm{100,max}^\mathrm{vq}$& 60 s\\ \hline
        Maximum admissible full activation time for FFR provision & \color{backgroundcolor2}$T_\mathrm{a,max}^\mathrm{ffr}$& 2 s\\
        Minimum support duration time for FFR provision & \color{backgroundcolor2}$T_\mathrm{d,min}^\mathrm{ffr}$& 8 s\\ 
        Minimum return-to-recovery time after FFR provision &\color{backgroundcolor2}$T_\mathrm{r,min}^\mathrm{ffr}$&10 s \\
        Overdelivery exceedance factor during FFR provision &\color{backgroundcolor2}$x_\mathrm{peak}^\mathrm{ffr}$&1.3 \\ \hline
        Active power droop gain for FCR provision &$D_\mathrm{p}$& 0.06\\
        Active power correlation factor for FFR provision &$K_\mathrm{p}$& 0.04\\
        Reactive power droop gain for voltage control &$D_\mathrm{q}$& 0.06\\
     \bottomrule
    \end{tabular}
    \vspace{-3mm}
     \label{tab:grid_code_parameters}
\end{table}
\renewcommand{\arraystretch}{1} \normalsize

\subsubsection*{Example 4 --- Superimposed FFR-FCR Provision} Although the FCR and FFR time-domain capability curves are typically specified separately as in \cref{fig:fcr_gridcode,fig:ffr_gridcode}, a single reserve unit can also provide both frequency regulation services simultaneously. However, depending on the future market situation, one might only be able to participate with the same amount of active power at one reserve market at a time (see, e.g., \cite{oyj2021technical}), i.e., the power sold to the FFR market cannot be sold to the FCR market. In respect thereof, a simple and pragmatic solution is to superimpose the FFR and FCR time-domain capability curves, where the time and capacity parameters can be maintained. A graphical illustration is exemplarily given in \cref{fig:ffr_fcr_superimposed_gridcode}, where the FFR capacity $|\Delta p_\mathrm{ffr}^\mathrm{n}|$ is larger than the FCR capacity $|\Delta p_\mathrm{fcr}^\mathrm{n}|$. However, also equivalent or reversed capacity sizes can be superimposed. For the superimposed FFR-FCR time-domain curve in \cref{fig:ffr_fcr_superimposed_gridcode}, we extend the \color{magenta} device-level \color{black} requirements in \eqref{eq:grid_code_req_fcr} and \eqref{eq:grid_code_req_ffr} accordingly as
\begin{subequations}\label{eq:grid_code_req_ffr_fcr_superimposed}
\begin{align}\label{eq:grid_code_req_ffr_fcr_superimposed1}
\tfrac{|\Delta p_\mathrm{fcr}^\mathrm{n}|}{\left(t_\mathrm{a}^\mathrm{fcr}-t_\mathrm{i}^\mathrm{fcr}\right)}+\tfrac{|\Delta p_\mathrm{ffr}^\mathrm{n}|}{t_a^\mathrm{ffr}} &\leq \color{magenta}R_\mathrm{max}^\mathrm{p}\color{black}\\\label{eq:grid_code_req_ffr_fcr_superimposed2}
|\Delta p_\mathrm{fcr}^\mathrm{n}|+|\Delta p_\mathrm{ffr}^\mathrm{peak,n}|&\leq \color{magenta}M_\mathrm{max}^\mathrm{p}\color{black}.
    \end{align}
\end{subequations}

In addition to the requirements in \cref{eq:grid_code_req_fcr,eq:grid_code_req_reactive_power,eq:grid_code_req_ffr,eq:grid_code_req_ffr_fcr_superimposed}, we assume that the services-providing reserve unit is able to supply the normalized
FCR capacity $|\Delta p^\mathrm{n}_\mathrm{fcr}|$, the normalized FFR capacity $|\Delta p^\mathrm{n}_\mathrm{ffr}|$, and the normalized reactive power capacity $|\Delta q_{100}^\mathrm{n}|$, respectively, when scaled with a \textit{non-unity} frequency and voltage deviation $|\Delta f|$ and $|\Delta v|$ up to a certain threshold, while at the same time being able to satisfy the minimum \color{backgroundcolor2} grid-code \color{black} requirements. Typical parameter values for the \color{backgroundcolor2} grid-code \color{black} specifications are provided in \cref{tab:grid_code_parameters}. 
Beyond that, notice that the active and reactive power time-domain capability curves in \cref{fig:gridcode_examples} represent simplified schematic examples of grid-code capability curves, which in practice are often more complex and sophisticated, while deviating for different grid codes and system operators. Furthermore, depending on the grid code, the capability curves can be superimposed in various ways.

\section{From Grid-Code Specifications to Rational Transfer Functions}\label{sec:grid_code_2_tf}
Although the specification of the previous piece-wise linear time-domain curves in the grid-code examples is straightforward, their practical implementation in a converter-based generation system is not immediate, and no systematic methods have been developed yet. In particular, as we will demonstrate later in our numerical experiments in \cref{sec:case_studies}, the well-known droop and virtual inertia control may not be able to achieve the required grid-code specifications due to their fixed controller structure (therefore may not pass the grid-code compliance tests), or would require a cumbersome trial-and-error tuning by adding filters to approximately satisfy the grid-code and device-level requirements. In this regard, today's industrial practice how to implement the required piece-wise linear time-domain grid-code curves is usually very ad-hoc and highly customized, e.g., relying on methods similar to open-loop trajectory commands, varying-gain, or look-up table schemes.

In this work, we thus propose a more systematic and versatile approach for the practical implementation of piece-wise linear time-domain curves to provide dynamic ancillary services by converter-based generation systems, while ensuring grid-code and device-level requirements to be reliably satisfied. More specifically, we aim to translate generic piece-wise linear time-domain curves for active and reactive power injection ($\Delta p$ and $\Delta q$, respectively) after a frequency and voltage step change ($\Delta f$ and $\Delta v$, respectively) into a \textit{desired} rational $2\times 2$ transfer function matrix $T_\mathrm{des}(s,\alpha)$ in the frequency domain, i.e.,
\begin{align}\label{eq:Tdes}
    \vspace{-1cm}
    \begin{bmatrix}
        \Delta p(s)\\ \Delta q(s)
    \end{bmatrix} = 
    \underset{=\,-T_\mathrm{des}(s,\alpha)}{\underbrace{\begin{bmatrix}
        -T_\mathrm{des}^\mathrm{fp}(s,\alpha) & 0 \\ 0 & -T_\mathrm{des}^\mathrm{vq}(s,\alpha) 
    \end{bmatrix}}}
    \begin{bmatrix}
        \Delta f(s) \\ \Delta v(s)
    \end{bmatrix},
    \vspace{-1cm}
\end{align}
which specifies a decoupled grid-following frequency and voltage control behavior of a reserve unit via $T_\mathrm{des}^\mathrm{fp}(s,\alpha)$ and $T_\mathrm{des}^\mathrm{vq}(s,\alpha)$, respectively. These two transfer functions are parametric in the vector $\alpha$ that contains the time-domain curve related parameters to be selected by the reserve unit, while ensuring that the \color{backgroundcolor2} grid-code \color{black} and \color{magenta} device-level \color{black}requirements are satisfied. More specifically, when translating the previous grid-code example for the superimposed FFR-FCR provision in \cref{fig:ffr_fcr_superimposed_gridcode} into $T_\mathrm{des}^\mathrm{fp}(s,\alpha) = T_\mathrm{des,fcr}^\mathrm{fp}(s,\alpha)+T_\mathrm{des,ffr}^\mathrm{fp}(s,\alpha)$, as well as the example for the voltage control in \cref{fig:vq_gridcode} into $T_\mathrm{des}^\mathrm{vq}(s,\alpha)$, the parameter vector $\alpha$ would be assembled as 
\begin{align}\label{eq:alpha}
\hspace{-2mm}\alpha = \begin{bmatrix}t_\mathrm{i}^\mathrm{fcr}&
t_\mathrm{a}^\mathrm{fcr}&
t_\mathrm{90}^\mathrm{vq}&
t_\mathrm{100}^\mathrm{vq}&
t_\mathrm{a}^\mathrm{ffr}&
t_\mathrm{d}^\mathrm{ffr}&
t_\mathrm{r}^\mathrm{ffr}&
|\Delta p_\mathrm{ffr}^\mathrm{peak,n}|
\end{bmatrix}^\top\hspace{-3mm}, 
\end{align}
and has to satisfy the \color{backgroundcolor2} grid-code \color{black} and \color{magenta} device-level \color{black}requirements in \cref{eq:grid_code_req_fcr,eq:grid_code_req_reactive_power,eq:grid_code_req_ffr,eq:grid_code_req_ffr_fcr_superimposed}. In the following, we present in detail how to obtain the parametric frequency-domain transfer functions $T_\mathrm{des}^\mathrm{fp}(s,\alpha)$ and $T_\mathrm{des}^\mathrm{vq}(s,\alpha)$ from any piece-wise linear time-domain grid-code curves. Afterwards, in \cref{sec:tf_based_control}, we show how these transfer functions can be easily implemented in standard converter control architectures. 

\begin{remark}\fontdimen2\font=0.6ex
In this work, we consider the active and reactive power time-domain capability curves in response to a general frequency and voltage step change $|\Delta f|$ and $|\Delta v|$. However, in practice, grid codes often consider different piece-wise linear time-domain response patterns for different operating ranges of $|\Delta f|$ and $|\Delta v|$. Also deadband or saturation thresholds for a particular range of $|\Delta f|$ and $|\Delta v|$ are usually defined \cite{european2016commission}. Although we omit these features in this paper, they can be easily realized in the reserve unit's converter control setup. 
\end{remark}

\begin{remark}\fontdimen2\font=0.6ex
For the transfer function matrix in \cref{eq:Tdes}, we primarily consider small-signal changes of $\Delta f$ and $\Delta v$ during normal operating conditions as usual for dynamic ancillary services provision. Hence, our work does not encompass the exploration of grid-code specifications pertaining to fault scenarios.
\end{remark}

\subsection{Translating Piece-Wise Linear Time-Domain Grid-Code Curves into Rational Parametric Transfer Functions}\label{sec:curve_to_tf} 
\begin{figure}[t!]
    \centering
   
    \begin{subfigure}{0.292\textwidth}
        \centering
         \vspace{-1mm}\resizebox{0.91\textwidth}{!}{
\begin{tikzpicture}
\draw[-latex] (2.3,0.9) -- (2.3,3.1); 
\draw [-latex](2.2,1) -- (6.9,1);
\node [scale=0.9] at (1.95,3.15) {$y(t)$};
\node [scale=0.9] at (6.9,0.8) {$t$};

\draw[dashed] (2.5,2.8) node (v2) {} -- (2.5,1);

\draw [dashed](3,2.3) -- (3,1); 
\draw [dotted,black!60](3,2.3) -- (2.3,2.3);

\node [scale=0.9] at (2.55,0.75) {$t_1$};

\node [scale=0.9] at (3,0.75) {$t_2$};

\draw[dotted,black!60](2.5,2.8)  -- (2.3,2.8);

\draw [ultra thick, black!40](2.3,1) --  (2.5,2.8) -- (3,2.3) -- (4.2,2.1) node (v1) {}--(4.6,1.5) node (v3) {}--(5.2,1.8) node (v4) {}--(6.3,1.8) node (v7) {} ;

\draw [dotted,black!60] (4.2,2.1)-- (2.3,2.1); 
\draw [dotted,black!60] (4.6,1.5)-- (2.3,1.5); 
\draw [dotted,black!60](5.2,1.8)  -- (2.3,1.8); 

\node [scale=0.9] at (2.3,0.75) {$t_0$}; 
\draw [dashed](4.2,2.1) -- (4.2,1); 

\draw[dashed](4.6,1.5) -- (4.6,1); 
\draw[dashed](5.2,1.8)  -- (5.2,1); 

\draw [dashed](6.3,1.8) -- (6.3,1);
\node [scale=0.9] at (4.2,0.75) {$t_3$};

\node [scale=0.9] at (4.6,0.75) {$t_4$};
\node [scale=0.9] at (5.2,0.75) {$t_5$};
\node [scale=0.9] at (6.3,0.75) {$t_6$};
\node [scale=0.8,black!60]at (2.05,1.5) {$y_4$};
\node [scale=0.8,black!60]at (1.85,1.8) {$y_5, y_6$};
\node [scale=0.8,black!60]at (2.05,2.1) {$y_3$};
\node [scale=0.8,black!60]at (2.05,2.3) {$y_2$};
\node [scale=0.8,black!60] at (2.05,2.8) {$y_1$};
\node [scale=0.8,black!60] at (2.05,1) {$y_0$};
\end{tikzpicture}
}
        \vspace{-9mm}
        \caption{Piece-wise linear time-domain curve with similar shape as the FFR-FCR curve in \cref{fig:ffr_fcr_superimposed_gridcode}.}
        \label{fig:piecewise_linear_curve}
    \end{subfigure}
    \hspace{1mm}
 \begin{subfigure}{0.18\textwidth}
        \centering
         \vspace{-1mm}\resizebox{0.97\textwidth}{!}{
\begin{tikzpicture}
\draw[-latex] (2.3,0.9) -- (2.3,3.1); 
\draw [-latex](2.2,1) -- (4.9,1);
\node [scale=0.9] at (1.85,3.15) {$y_{ij}(t)$};
\node [scale=0.9] at (4.9,0.8) {$t$};
\draw [dotted,black!60](2.9,1.7) -- (2.2,1.7);

\node [scale=0.9] at (2.9,0.75) {$t_i$};

\node [scale=0.9]at (4.2,0.75) {$t_j$};

\draw[dotted,black!60](4.2,2.4)  -- (2.2,2.4);

\draw [ultra thick, black!40](2.3,1)--(2.9,1) --  (2.9,1.7) -- (4.2,2.4)--(4.2,1)--(4.7,1);

\node [scale=0.8,black!60] at (2.05,1.7) {$y_i$};
\node [scale=0.8,black!60] at (2.05,2.4) {$y_j$};

\end{tikzpicture}
}
        \vspace{-9.1mm}
        \caption{Linear curve segment.\color{white}das istein seer langes wort}
        \label{fig:curve segment}
    \end{subfigure}
    \caption{Normalized unit step response time-domain capability curve of a general grid-code specification with linear curve segments.}
\vspace{-0.2cm}
    \label{fig:gridcode2tf}
\end{figure}
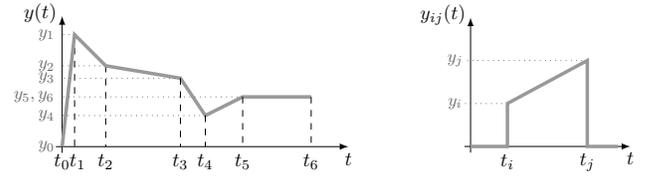

\fontdimen2\font=0.6ex
In this section, we present a general procedure on how to translate a piece-wise linear step-response capability curve which is specified in the time domain into a rational parametric transfer function in the frequency domain. Our approach is based on the assumption that the time-domain curve reflects a \textit{stable} step-response behavior $y(t)$ under a unit step input $u(t)=u_\mathrm{step}=1$. Moreover, each curve kink is assumed to be characterized by a time-capacity parameter pair $(t_i,y_i),\,i\in\mathbb{N}$, where the normalized capacity $y_i=K_iu_\mathrm{step}=K_i$ is scaled by the unit step input via some gain $K_i\in\mathbb{R}$ (cf. the droop gains introduced in \cref{sec:grid_code_curves}). For ease of translation, we consider a unit-step response ($u_\mathrm{step} = 1$). A general representation of a normalized piece-wise linear time-domain response curve with a similar shape as the one in \cref{fig:ffr_fcr_superimposed_gridcode} is illustrated in \cref{fig:piecewise_linear_curve}.

The procedure to obtain a rational transfer function representation of such a piece-wise linear time-domain response curve, consists of four steps: In a \textit{first} step, we decompose the overall piece-wise linear time-domain response curve $y(t)$ in \cref{fig:piecewise_linear_curve} into linear curve segments $y_{ij}(t),\,i,j\in\mathbb{N}$ as indicated in \cref{fig:curve segment}. Each unit step response curve segment $y_{ij}(t)$ is characterized by two time-capacity parameter pairs $(t_i,y_i)$ and $(t_j,y_j)$, where $j>i$ and thus $t_j>t_i$, such that the curve segment can be described in the time domain as 
\begin{align}\label{eq:curve_segment_t_domain}
y_{ij}(t) = \begin{cases}\tfrac{y_j-y_i}{t_j-t_i}t+y_i & t_i\leq t\leq t_j\\
0 & \text{else},
\end{cases}
\end{align}
where we define $d=\tfrac{y_j-y_i}{t_j-t_i}$ as the slope of the curve segment. Obviously, depending on the capacities $y_i$ and $y_j$, the curve segment is either increasing for $y_j>y_i$ (i.e., $d >0$), decreasing for $y_j<y_i$ (i.e., $d<0$), or flat for $y_j=y_i$ (i.e., $d=0$).

Next, in a \textit{second} step, we apply the Laplace transformation to each time-domain curve segment $y_{ij}(t)$ in \eqref{eq:curve_segment_t_domain} and obtain the unit step response of a curve segment in the $s$-domain as
\begin{align}\label{eq:curve_segment_s_domain_step_response}
    Y_{ij}(s) = \left( \tfrac{y_i}{s}+\tfrac{d}{s^2} \right ) e^{-t_is}- \left( \tfrac{y_j}{s}+\tfrac{d}{s^2} \right ) e^{-t_js},
\end{align}
where we have used the time-shift property of the Laplace transformation \cite{franklin2002feedback}. The actual transfer function or impulse response of the curve segment can be computed by multiplying the unit step response in \eqref{eq:curve_segment_s_domain_step_response} by $s$ as $T_{ij}^\mathrm{uy}(s)=sY_{ij}(s)$, i.e., 
\begin{align}\label{eq:curve_segment_s_domain_tf}
    T_{ij}^\mathrm{uy}(s) = \left( y_i+\tfrac{d}{s} \right ) e^{-t_is}- \left( y_j+\tfrac{d}{s} \right ) e^{-t_js},
\end{align}
which corresponds to a non-rational transfer function. From a practical point of view, however, such a non-rational transfer function is not easily interpretable and implementable, e.g., when using standard discretization and micro-controller interfaces. On top of that, most methods for analysis and synthesis of control systems, are developed for rational transfer functions \cite{franklin2002feedback} (e.g., to determine stability, assess passivity, Nyquist and Root Locus methods, robust and optimal control methods).

In this regard, as a \textit{third} step, we need to approximate \eqref{eq:curve_segment_s_domain_tf} by a rational transfer function. To do so, we apply the commonly used \textit{Padé}-approximation with numerator and denominator degree $n\in\mathbb{N}$ to every exponential as \cite{hanta2009rational} 
    \begin{align}\label{eq:pade}
        e^{-t_is}\approx \left(1-\tfrac{t_i}{2n}s\right)^n / \left(1+\tfrac{t_i}{2n}s\right)^n, 
    \end{align}
which approximates \eqref{eq:curve_segment_s_domain_tf} as a rational transfer function, i.e.,
\begin{align}\label{eq:curve_segment_s_domain_rational_tf}
    \hspace{-1mm} T_{ij}^\mathrm{uy}(s) \hspace{-0.5mm}\approx\hspace{-0.5mm} \left( y_i \hspace{-0.5mm}+ \hspace{-0.5mm}\tfrac{d}{s} \right ) \hspace{-0.5mm} \tfrac{\left(1-\tfrac{t_i}{2n}s\right)^n}{\left(1+\tfrac{t_i}{2n}s\right)^n} \hspace{-0.5mm}- \hspace{-0.5mm} \left( y_j \hspace{-0.5mm}+ \hspace{-0.5mm}\tfrac{d}{s} \right ) \hspace{-0.5mm}\tfrac{\left(1-\tfrac{t_j}{2n}s\right)^n}{\left(1+\tfrac{t_j}{2n}s\right)^n}.
\end{align}
Alternatively, one might also resort to more general types of Padé-approximations or other rational series expansions.

\begin{figure}[t!]
    \centering
    \hspace{-5mm}
\begin{subfigure}{0.23\textwidth}
\centering
\hspace{-1mm}
    \scalebox{0.46}{\includegraphics[]{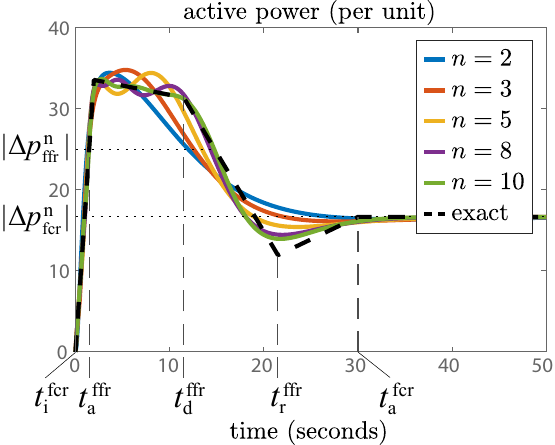}}
    \vspace{-0.5cm}
    \caption{Approximation of the normalized piece-wise linear time-domain curve for FFR-FCR provision in \cref{fig:ffr_fcr_superimposed_gridcode}.}
    \label{fig:ffr_fcr_pade}
\end{subfigure}
\hspace{1.5mm}
\begin{subfigure}{0.23\textwidth}
\centering
    \scalebox{0.46}{\includegraphics[]{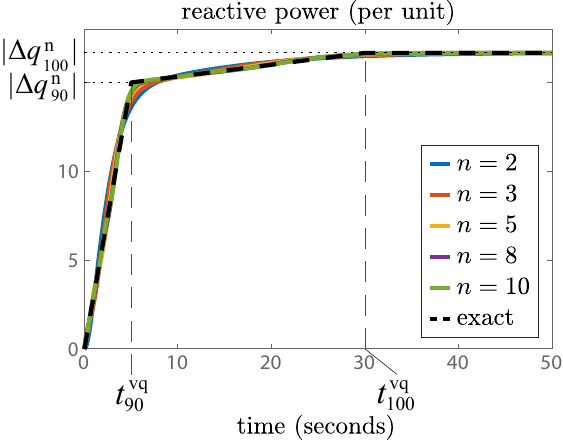}}
    \vspace{-0.5cm}
    \caption{Approximation of the normalized piece-wise linear time-domain curve for voltage control in \cref{fig:vq_gridcode}.}
    \label{fig:vq_pade}
\end{subfigure}
    \caption{Unit step response of the rational transfer functions (a) $T_\mathrm{des}^\mathrm{fp}(s,\alpha)$ and (b) $T_\mathrm{des}^\mathrm{vq}(s,\alpha)$ for different orders $n$ of the Padé-approximation.}
     \vspace{-4mm}
    \label{fig:pade_accuracy_examples}
\end{figure}

Finally, in a \textit{fourth} step, the overall rational transfer function of the normalized piece-wise linear time-domain unit step response capability curve in \cref{fig:piecewise_linear_curve} can be established as the sum of the rational transfer functions of the linear curve segments in \cref{fig:curve segment}, i.e., $T_\mathrm{des}^\mathrm{uy}(s)=\textstyle\sum T_{ij}^\mathrm{uy}(s)$. Notice that the obtained transfer function $T_\mathrm{des}^\mathrm{uy}(s)$ is parametric in the time and capacity parameters $t_i$ and $y_i$, respectively, which are either directly fixed by the grid-code or can be appropriately selected by the reserve unit. In accordance with the notation introduced in \eqref{eq:Tdes}, one might therefore equivalently write $T_\mathrm{des}^\mathrm{uy}(s)=T_\mathrm{des}^\mathrm{uy}(s,\alpha)$ with $\alpha=[..., t_i, ..., y_i,...]^\top$, to indicate the parametric dependence. 

Coming back to the grid-code examples in \cref{sec:grid_code_curves}, we translate the superimposed FFR-FCR time-domain capability curve in \cref{fig:ffr_fcr_superimposed_gridcode} and the reactive power time-domain capability curve in \cref{fig:vq_gridcode} into rational parametric transfer functions $T_\mathrm{des}^\mathrm{fp}(s,\alpha) = T_\mathrm{des,fcr}^\mathrm{fp}(s,\alpha) + T_\mathrm{des,ffr}^\mathrm{fp}(s,\alpha)$ and $T_\mathrm{des}^\mathrm{vq}(s,\alpha)$, respectively. For a feasible choice of $\alpha$ in \eqref{eq:alpha} satisfying \cref{eq:grid_code_req_fcr,eq:grid_code_req_reactive_power,eq:grid_code_req_ffr,eq:grid_code_req_ffr_fcr_superimposed} with the grid-code specifications in \cref{tab:grid_code_parameters}, and for a sufficiently flexible reserve unit as in \cref{sec:case_studies}, the unit step response of the latter transfer functions for different orders $n$ of the Padé-approximation is shown in \cref{fig:pade_accuracy_examples}. We can see how a higher order improves the approximation accuracy of the piece-wise linear time-domain curve. However, since a too large order typically becomes numerically intractable, and, additionally, might not be realizable in a practical power converter control architecture, we recommend to choose orders $n\leq 10$. To gain some insights on the structure of $T_\mathrm{des}(s,\alpha)$, the transfer functions for $n=2$ in \cref{fig:pade_accuracy_examples} are exemplarily given as
\begin{align}\nonumber
    T_\mathrm{des,fcr}^\mathrm{fp}(s,\alpha) &=\tfrac{0.2963}{s^2 + 0.2667 s + 0.01778} \\ \nonumber
    T_\mathrm{des,ffr}^\mathrm{fp}(s,\alpha) &=\tfrac{143.7 s^4 + 154.6 s^3 + 59.75 s^2 + 7.599 s}{s^6 + 5.17 s^5 + 9 s^4 + 6.26 s^3 + 2.03 s^2 + 0.3077 s + 0.0176} \\\label{eq:Tdes_example_structure}
    T_\mathrm{des}^\mathrm{vq}(s,\alpha) &= \tfrac{9.422 s^2 + 2.56 s + 0.1897}{s^4 + 1.867 s^3 + 1.084 s^2 + 0.1991 s + 0.01137}.
\end{align}

Finally, our approach to translate a piece-wise linear time-domain curve into a rational transfer function always results in a stable transfer function $T_\mathrm{des}^\mathrm{uy}(s,\alpha)$ for any order $n$ of the Padé-approximation. Indeed, the rational transfer function $T_{ij}^\mathrm{uy}(s,\alpha)$ in \eqref{eq:curve_segment_s_domain_rational_tf} of one curve segment can be rewritten as in \eqref{eq:stability_proof}, where the stability of the last term follows from the fact that the numerator has a zero at $s=0$ which cancels with the pole at $s=0$ in the denominator, i.e., 
\begin{align}\nonumber
    T_{ij}^\mathrm{uy}(s,\alpha) &\approx  \left( y_is + d \right )  \tfrac{\left(1-\tfrac{t_i}{2n}s\right)^n}{s\left(1+\tfrac{t_i}{2n}s\right)^n}- \left( y_j s+ d \right ) \tfrac{\left(1-\tfrac{t_j}{2n}s\right)^n}{s\left(1+\tfrac{t_j}{2n}s\right)^n}\\
\label{eq:stability_proof}
    \color{white}T_{ij}^\mathrm{uy}(s,\alpha) \color{black}&= \underset{\text{stable}}{\underbrace{y_i \tfrac{\left(1-\tfrac{t_i}{2n}s\right)^n}{\left(1+\tfrac{t_i}{2n}s\right)^n}}} - \underset{\text{stable}}{\underbrace{y_j\tfrac{\left(1-\tfrac{t_j}{2n}s\right)^n}{\left(1+\tfrac{t_j}{2n}s\right)^n}}}\\\nonumber
    &\quad +\underset{\text{stable}}{\underbrace{d \tfrac{\left( 1-\tfrac{t_i}{2n}s\right)^n\left( 1+\tfrac{t_j}{2n}s\right)^n-\left( 1-\tfrac{t_j}{2n}s\right)^n\left( 1+\tfrac{t_i}{2n}s\right)^n}{s \left( 1+\tfrac{t_i}{2n}s\right)^n\left( 1+\tfrac{t_j}{2n}s\right)^n}.}} 
\end{align}
We can thus conclude that the rational transfer function $T_{ij}^\mathrm{uy}(s,\alpha)$ of one curve segment is stable, and with this the sum $T_\mathrm{des}^\mathrm{uy}(s,\alpha)=\textstyle\sum T_{ij}^\mathrm{uy}(s,\alpha)$. 

Next, we provide insights on how to select the time-domain-curve related parameters $\alpha$. For the sake of consistency, we stick to the grid-code examples for frequency and voltage control in \cref{fig:ffr_fcr_superimposed_gridcode,fig:vq_gridcode} for the remainder of this paper.

\subsection{Parameter Selection}\label{sec:parameter_selection}
\begin{figure}[t!]
    \centering
\begin{subfigure}{0.5\textwidth}
\centering
\hspace{-1mm}
    \scalebox{0.47}{\includegraphics[]{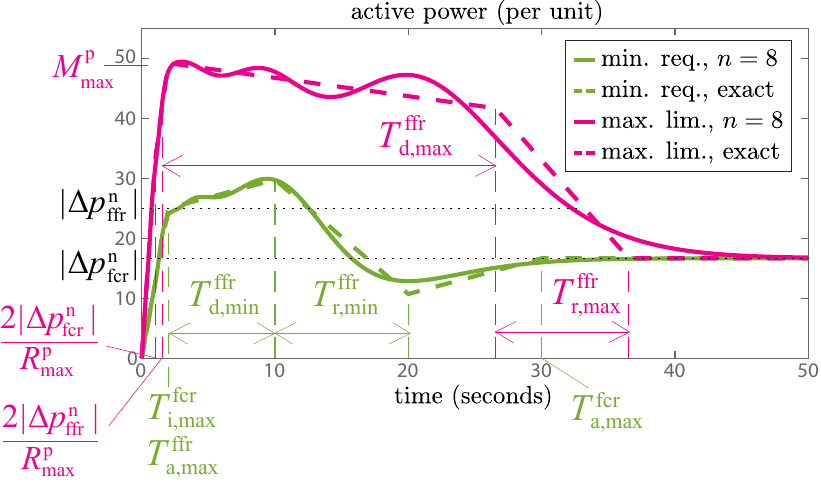}}
    \vspace{-0.1cm}
    \caption{Approximation of the normalized piece-wise linear time-domain curve for FFR-FCR provision in \cref{fig:ffr_fcr_superimposed_gridcode}.}
    \label{fig:ffr_fcr_boundary}
\end{subfigure}
\begin{subfigure}{0.5\textwidth}
\centering
\vspace{0.2cm}
    \scalebox{0.47}{\includegraphics[]{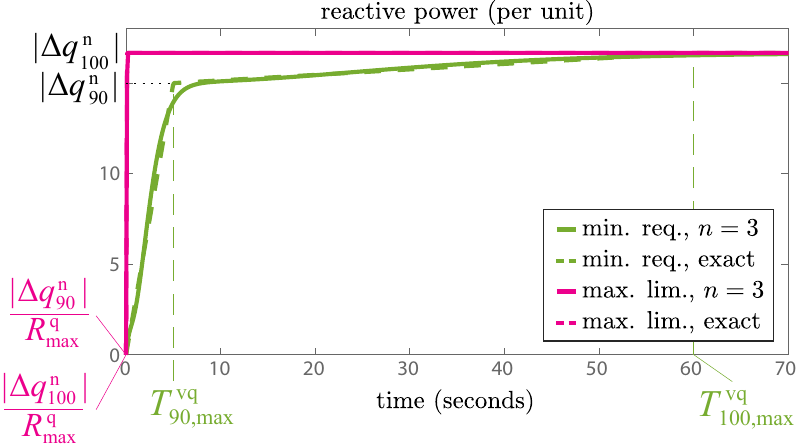}}
    \vspace{-0.1cm}
    \caption{Approximation of the normalized piece-wise linear time-domain reactive power curve for voltage control in \cref{fig:vq_gridcode}.}
    \label{fig:vq_boundary}
\end{subfigure}
    \caption{Unit step responses of (a) $T_\mathrm{des}^\mathrm{fp}(s,\alpha)$ and (b) $T_\mathrm{des}^\mathrm{vq}(s,\alpha)$ for the two boundary scenarios of the parameter choice $\alpha$: satisfying \color{backgroundcolor2} minimum grid-code requirements \color{black} vs. exploiting \color{magenta} maximum device-level limitations\color{black}.}
     \vspace{-4mm}
    \label{fig:boundary_case_comparison}
\end{figure}

\fontdimen2\font=0.6ex
Given the \color{backgroundcolor2} grid-code \color{black} and \color{magenta} device-level \color{black} requirements in \cref{eq:grid_code_req_fcr,eq:grid_code_req_reactive_power,eq:grid_code_req_ffr,eq:grid_code_req_ffr_fcr_superimposed}, it follows that the time-domain-curve related parameters $\alpha$ in \eqref{eq:alpha} allow for a set of feasible rational transfer functions $T_\mathrm{des}(s,\alpha)$ in \eqref{eq:Tdes} to provide a decoupled frequency and voltage regulation. Among the various options on how to select the parameters $\alpha$, we can consider two boundary scenarios: One, where the parameters $\alpha$ satisfy the \color{backgroundcolor2} minimum grid-code requirements\color{black}, and one, where the parameters $\alpha$ are selected such that the \color{magenta} maximum device-level limitations \color{black} are exploited. In case of the former, we select the parameters $\alpha$ as
\begin{align}\label{eq:min_grid_code_req}
    \begin{split}
        \hspace{-2mm}&t_\mathrm{i}^\mathrm{fcr} = \color{backgroundcolor2} T_\mathrm{i,max}^\mathrm{fcr} \color{black}\quad\quad t_\mathrm{90}^\mathrm{vq} = \color{backgroundcolor2} T_\mathrm{90,max}^\mathrm{vq} \color{black}\quad\quad
        t_\mathrm{a}^\mathrm{ffr} = \color{backgroundcolor2} T_\mathrm{a,max}^\mathrm{ffr} \color{black}\\
        \hspace{-2mm}&t_\mathrm{a}^\mathrm{fcr} = \color{backgroundcolor2} T_\mathrm{a,max}^\mathrm{fcr}\color{black}\quad\,\,\,\,\,
        t_\mathrm{100}^\mathrm{vq} = \color{backgroundcolor2} T_\mathrm{100,max}^\mathrm{vq}\color{black}\,\,\,\,\,\,\,\hspace{0.15mm}
        t_\mathrm{d}^\mathrm{ffr}-t_\mathrm{a}^\mathrm{ffr} = \color{backgroundcolor2} T_\mathrm{d,min}^\mathrm{ffr}\color{black}\\
        \hspace{-2mm}&\quad\quad\quad\quad\quad\quad\quad\quad\quad\quad\quad\quad\quad\quad\quad t_\mathrm{r}^\mathrm{ffr}-t_\mathrm{d}^\mathrm{ffr} = \color{backgroundcolor2} T_\mathrm{r,min}^\mathrm{ffr}\color{black}\\
        \hspace{-2mm}&\quad\quad\quad\quad\quad\quad\quad\quad\quad\quad\quad\quad\quad\quad\quad |\Delta p_\mathrm{ffr}^\mathrm{peak,n}| = \color{backgroundcolor2} |\Delta p_\mathrm{ffr}^\mathrm{n}|\color{black},\hspace{-4mm}
    \end{split}
\end{align}
where typical values for the \color{backgroundcolor2} grid-code \color{black} specifications are given in \cref{tab:grid_code_parameters}. On the other hand, for the latter, we specify
\begin{align}\label{eq:max_dev_level_req}
    \begin{split}
        \hspace{-2mm}&t_\mathrm{i}^\mathrm{fcr} = \color{magenta} 0 \color{black}\quad\quad\quad\quad\,\, t_\mathrm{90}^\mathrm{vq} = \tfrac{|\Delta q_\mathrm{90}^\mathrm{n}|}{\color{magenta} R^\mathrm{q}_\mathrm{max}\color{black}}\quad\quad
        t_\mathrm{a}^\mathrm{ffr} = \tfrac{2|\Delta p_\mathrm{ffr}^\mathrm{n}|}{\color{magenta} R^\mathrm{p}_\mathrm{max}\color{black}}\\
        \hspace{-2mm}&t_\mathrm{a}^\mathrm{fcr} = \tfrac{2|\Delta p_\mathrm{fcr}^\mathrm{n}|}{\color{magenta} R^\mathrm{p}_\mathrm{max}\color{black}}\quad\,\,\,\,\,
        t_\mathrm{100}^\mathrm{vq} = \tfrac{|\Delta q_\mathrm{100}^\mathrm{n}|}{\color{magenta} R^\mathrm{q}_\mathrm{max}\color{black}}\,\,\,\,\,\,\,\,
        t_\mathrm{d}^\mathrm{ffr}-t_\mathrm{a}^\mathrm{ffr} = \color{magenta} T_\mathrm{d,max}^\mathrm{ffr}\color{black}\\
        \hspace{-2mm}&\quad\quad\quad\quad\quad\quad\quad\quad\quad\quad\quad\quad\quad\quad\,\,\,\, t_\mathrm{r}^\mathrm{ffr}-t_\mathrm{d}^\mathrm{ffr} = \color{magenta} T_\mathrm{r,max}^\mathrm{ffr}\color{black}\\
        \hspace{-2mm}&\quad\quad\quad\quad\quad\quad\quad\quad\quad \quad\quad\quad\quad\quad\,\,\,\,\scriptstyle|\Delta p_\mathrm{ffr}^\mathrm{peak,n}|+|\Delta p_\mathrm{fcr}^\mathrm{n}| = \color{magenta} M_\mathrm{max}^\mathrm{p} \color{black},\normalsize\hspace{-10mm}
    \end{split}
\end{align}
where $|\Delta p_\mathrm{ffr}^\mathrm{peak,n}|\leq \color{backgroundcolor2}x_\mathrm{peak}^\mathrm{ffr}\cdot |\Delta p_\mathrm{ffr}^\mathrm{n}|\color{black}$, and \color{magenta} $R_\mathrm{max}^\mathrm{p}$, $R_\mathrm{max}^\mathrm{q}$, $T_\mathrm{d,max}^\mathrm{ffr}$, $T_\mathrm{r,max}^\mathrm{ffr}$, \color{black} and \color{magenta} $M_\mathrm{max}^\mathrm{p}$ \color{black} are \color{magenta} device-level \color{black} specifications of the ancillary services providing reserve unit. In this regard, for an exemplary reserve unit as in \cref{sec:case_studies}, we obtain the unit step responses of the two boundary scenarios for the superimposed FFR-FCR provision, as well as the voltage control as in \cref{fig:boundary_case_comparison}. The first scenario which encodes the \color{backgroundcolor2} minimum grid-code requirements\color{black}, offers a cheap and robust (i.e., with smaller gain) version to provide dynamic ancillary services. In contrast, the second scenario encodes a response behavior that \color{magenta} reaches the limit of the reserve unit\color{black}, but offers the possibility to improve the overall frequency response of the grid. All remaining feasible choices of $\alpha$ are somewhere in between these two boundary scenarios (albeit not strictly between the two curves).

Ideally, the choice of $\alpha$ should be based on some stability and/or performance criteria when considering $T_\mathrm{des}(s,\alpha)$ in closed-loop with the grid. In other words, system-level stability should be ensured by properly specifying the desired ancillary service transfer function $T_\mathrm{des}(s,\alpha)$ given the flexibility of grid-code and device-level requirements. Namely, future ancillary service markets might provide some incentives on providing grid services beyond the \color{backgroundcolor2} minimum grid-code requirements\color{black}, i.e., by rewarding additional effort in improving the overall grid response behavior, while taking local device-level limitations into account. This might also motivate for a parameter-varying ancillary services specification $T_\mathrm{des}(s,\alpha)$ which is adapted online during different grid conditions. Making these considerations more concrete has been elaborated in our recent work in \cite{haberle2024optimal} and is part of our ongoing research.
\renewcommand{\arraystretch}{1.2}
\begin{table}[b!]\scriptsize
    \centering
    \vspace{-3mm}
     \setlength{\tabcolsep}{1mm}
           \caption{Parameters for the converter-interfaced generation system.}
               \begin{tabular}{c||c|c}
     \toprule
          Parameter & Symbol & Value  \\ \hline
          Voltage, power \& frequency base value & $V_\mathrm{b}$, $S_\mathrm{b}$, $f_\mathrm{b}$ & 1 kV, 10 MVA, 50 Hz\\
            DC-link capacitor& $C_\mathrm{dc}$ & 0.24 p.u.\\
            $RL$-filter components & $L_\mathrm{f}$, $R_\mathrm{f}$ & 0.1 p.u., 0.01 p.u. \\
            DC-source time-constant & $\tau_\mathrm{dc}$ & 0.5 s\\ \hline
            PLL control gains & $k_\mathrm{p}^\mathrm{pll}$, $k_\mathrm{i}^\mathrm{pll}$& 0.57, 10.19 \\
            Inner current control gains &$k_\mathrm{p}^\mathrm{i}$, $k_\mathrm{i}^\mathrm{i}$& 0.32, 10 \\
            DC-voltage control gains &$k_\mathrm{p}^\mathrm{dc}$, $k_\mathrm{i}^\mathrm{dc}$& 200, 1200\\
            Reactive power control gains &$k_\mathrm{p}^\mathrm{q}$, $k_\mathrm{i}^\mathrm{q}$& 3, 100 \\
            Active power control gains &$k_\mathrm{p}^\mathrm{p}$, $k_\mathrm{i}^\mathrm{p}$& 20, 100\\ \hline
            Normalized max. active power ramping rate & \color{magenta} $R_\mathrm{max}^\mathrm{p}$& 32.56 p.u./s\\
            Normalized max. reactive power ramping rate & \color{magenta} $R_\mathrm{max}^\mathrm{q}$& 150 p.u./s\\
            Max. FFR support duration & \color{magenta} $T_\mathrm{d,max}^\mathrm{ffr}$ & 25 s\\
            Max. return-to-recovery time & \color{magenta} $T_\mathrm{r,max}^\mathrm{ffr}$& 10 s\\
            Normalized max. active power peak capacity &\color{magenta} $M_\mathrm{max}^\mathrm{p}$ & 49.167 p.u.\\
     \bottomrule
    \end{tabular}
     \label{tab:converter_parameters}
\end{table}
\renewcommand{\arraystretch}{1} \normalsize

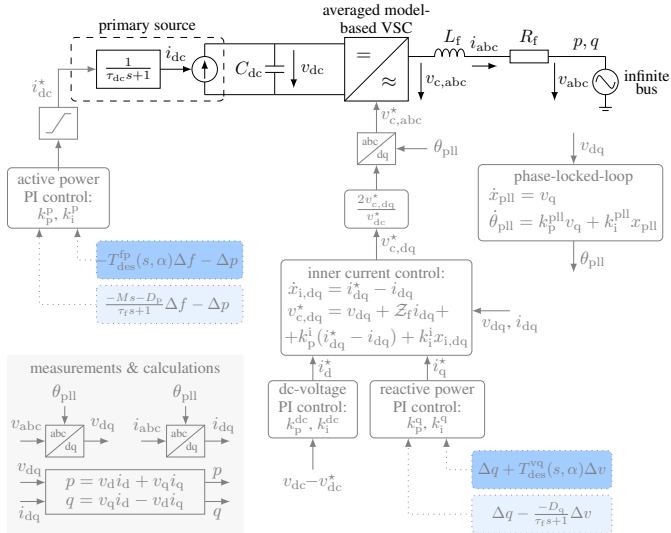
\begin{figure}[b!]
    \centering
        \vspace{-4mm}
     \usetikzlibrary{circuits.ee.IEC}
\resizebox {0.5\textwidth}{!}{
\begin{tikzpicture}[circuit ee IEC, scale =1, every node/.style={scale = 0.8}]
\draw [fill=black!3,color=black!3] (-6.8,-6) rectangle (-3.1,-3.2);
\draw  (-1.5,1.8) rectangle (-0.5,0.8);
\node at (-2,1.3) {$v_\mathrm{dc}$};
\draw (-1.5,0.8) -- (-0.5,1.8);
\node [scale=1.1] at (-1.2,1.5) {$=$};
\node [scale=1.1] at (-0.8,1.1) {$\approx$};
%\node [scale=0.9] at (-1,2.45) {averaged};
\node [scale=0.9] at (-1,2.21) {averaged model-};
\draw (-1.5,1.7) -- (-2.6,1.7) node (v1) {}; 
\draw (-1.5,0.9) -- (-2.6,0.9) node (v2) {}; 
\draw (-2.6,1.7) node (v3) {} -- (-2.6,1.35); 
\draw (-2.6,0.9) node (v4) {} -- (-2.6,1.25); 
\draw (-2.75,1.35) -- (-2.45,1.35); 
\draw (-2.75,1.25) -- (-2.45,1.25); 
\draw [-latex](-2.3,1.6) -- (-2.3,1); 
\draw(-2.6,1.7) -- (-3.7,1.7) node (v5) {}; 
\draw (-2.6,0.9) -- (-3.7,0.9) node (v6) {}; 
\draw  (-3.7,1.3) ellipse (0.2 and 0.2); 
\draw (-3.7,1.5) -- (-3.7,1.7) ; 
\draw (-3.7,1.1) -- (-3.7,0.9); 
\draw[-latex] (-3.7,1.15) -- (-3.7,1.45); 
\draw [-latex](-4.4,1.3)-- (-3.9,1.3);
\draw  (-4.4,1.6) rectangle (-5.4,1);
\node at (-4.15,1.55) {$i_\mathrm{dc}$};
\node at (-4.9,1.3) {$\frac{1}{\tau_\mathrm{dc}s+1}$}; 
\node [black!50] at (-6.25,1) {$i_\mathrm{dc}^\star$}; 

\draw [-latex,black!50] (-6,0.7) -- (-6,1.3) --(-5.4,1.3);
\draw [dashed,rounded corners = 2] (-3.4,0.8) rectangle (-5.8,1.8);
\node [scale=0.9] at (-4.6,1.97) {primary source};
\draw [black!50] (-6.3,0.7) rectangle (-5.7,0.2); 
\draw[black!50] (-6.2,0.3) -- (-6.1,0.3) -- (-5.9,0.6) -- (-5.8,0.6); 
\draw[-latex,black!50] (-6,-0.3) -- (-6,0.2); 

\draw [rounded corners = 2,black!50] (-6.8,-0.25) rectangle (-5.2,-1.2);
\node [scale=0.9,black!50] at (-6,-0.45) {active power};
\draw [dotted,-latex,black!50](-5.4,-1.7) -- (-5.7,-1.7) -- (-5.7,-1.2); 
\draw [dotted,-latex,black!50](-5.4,-2.4) -- (-6.3,-2.4) -- (-6.3,-1.2); 
\draw [dotted,black!50,fill=hydro!40,rounded corners = 2] (-5.4,-1.4) rectangle (-3.1,-2); 
\draw [dotted,black!50,fill=hydro!10,rounded corners = 2](-5.4,-2.1) rectangle (-3.1,-2.7);
\node [scale=0.9,black!50] at (-4.3,-1.7) {$-T_\mathrm{des}^\mathrm{fp}(s,\alpha)\Delta f-\Delta p$};
\node [scale=0.9,black!50] at (-4.3,-2.4) {$\tfrac{-Ms-D_\mathrm{p}}{\tau_\mathrm{f}s+1}\Delta f - \Delta p$};
\draw[-latex,black!50] (-1,0.3) -- (-1,0.8);

draw[-latex] (5.3,-2.7) -- (5.3,-2.3);
\node [black!50]at (-2,-5.2) {$v_\mathrm{dc}\hspace{-1.1mm}-\hspace{-0.8mm}v_\mathrm{dc}^\star$};
\draw [dotted,-latex,black!50] (0.5,-5) -- (0.1,-5) -- (0.1,-4.5); 
\draw  [dotted,-latex,black!50](0.5,-5.7) -- (-0.5,-5.7) -- (-0.5,-4.5);

\draw  [dotted,black!50,fill=hydro!40,rounded corners = 2](0.5,-4.7) rectangle (2.7,-5.3);
\draw  [dotted,black!50,fill=hydro!10,rounded corners = 2] (0.5,-5.4) rectangle (2.7,-6);
\node [scale=0.9,black!50] at (1.6,-5) {$\Delta q+T_\mathrm{des}^\mathrm{vq}(s,\alpha)\Delta v$};
\node [scale=0.9,black!50] at (1.6,-5.7) {$\Delta q-\tfrac{-D_\mathrm{q}}{\tau_\mathrm{f}s+1}\Delta v$};
\draw (-0.5,1.5) to [inductor={yscale=1.2,xscale=0.6}] (0.8,1.5) node (v9) {};
\draw  (1.1,1.6) rectangle (1.7,1.4); 
\draw (1.7,1.5) -- (2.6,1.5) node (v7) {}; 
\draw [-latex](1.8,1.4) -- (1.8,0.8);
\node at (-3,1.3) {$C_\mathrm{dc}$};

\node at (2.1,1.1) {$v_\mathrm{abc}$};

\draw  (2.6,1.1) node (v8) {} ellipse (0.2 and 0.2); 
\draw (2.6,0.9) -- (2.6,0.7); 
\draw (2.6,1.3) -- (2.6,1.5) ; 
\draw (2.5,0.7) -- (2.7,0.7);
\draw (2.55,0.65) -- (2.65,0.65); 
\node at (0.2,1.8) {$L_\mathrm{f}$};
\node at (1.4,1.8) {$R_\mathrm{f}$};
\draw  plot[smooth, tension=.7] coordinates {(2.45,1.1) (2.5,1.2) (v8) (2.7,1) (2.75,1.1)};

\node at (2.3,1.7) {$p,q$};
\node [scale=0.9] at (3.25,1.2) {infinite};
\node [scale=0.9] at (3.25,0.99) {bus};
\node [scale=0.9,black!50] at (-6,-1) { $k_\mathrm{p}^\mathrm{p}$, $k_\mathrm{i}^\mathrm{p}$};
\node [scale = 0.9,black!50] at (-6,-0.7) {PI control:};

\draw [-latex](-0.3,1.4) -- (-0.3,0.8);i
\node at (0.1,1.1) {$v_\mathrm{c,abc}$};

\draw [rounded corners = 2,black!50] (-1.5,-0.6) rectangle (-0.5,-1.2);
\node [black!50]at (-1,-0.9) {$\tfrac{2v_\mathrm{c,dq}^\star}{v_\mathrm{dc}^\star}$};
\draw[-latex,black!50] (-1,-0.6) -- (-1,-0.2);
\node[black!50] at (-0.6,0.55) {$v_\mathrm{c,abc}^\star$};

\draw [rounded corners = 2,black!50] (-2.5,-1.7) rectangle (0.5,-3.2);
\draw [-latex,black!50](-1,-1.7) -- (-1,-1.2);
\node [scale=0.9,black!50]at (-1,-1.9) {inner current control:};
\draw [black!50](-6.2,-5) rectangle (-3.7,-5.7);
\node [scale =0.9,black!50] at (-4.95,-3.4) {measurements \& calculations};
\node [black!50] at (-4.95,-5.2) {$p = v_\mathrm{d}i_\mathrm{d}+v_\mathrm{q}i_\mathrm{q}$};
\node [black!50] at (-4.95,-5.5) {$q=v_\mathrm{q}i_\mathrm{d}-v_\mathrm{d}i_\mathrm{q}$};
\draw [-latex,black!50](-6.6,-5.2) -- (-6.2,-5.2); 
\draw [-latex,black!50](-6.6,-5.5) -- (-6.2,-5.5); 
\node  [black!50] at (-6.45,-5) {$v_\mathrm{dq}$};
\node [black!50] at (-6.45,-5.7) {$i_\mathrm{dq}$};
\draw [-latex,black!50] (-3.7,-5.2) -- (-3.3,-5.2); 
\draw  [-latex,black!50](-3.7,-5.5) -- (-3.3,-5.5);
\node [black!50] at (-3.5,-5.05) {$p$};
\node [black!50] at (-3.5,-5.7) {$q$};
\node [black!50]at (-1.4,-2.2) {$\dot{x}_\mathrm{i,dq}=i_\mathrm{dq}^\star-i_\mathrm{dq}$};
\node [black!50]at (-1.125,-2.55) {$v_\mathrm{c,dq}^\star=v_\mathrm{dq}+\mathcal{Z}_\mathrm{f}i_\mathrm{dq}+$};
\node [black!50]at (-1,-2.94) {$+k_\mathrm{p}^\mathrm{i}(i_\mathrm{dq}^\star-i_\mathrm{dq})+k_\mathrm{i}^\mathrm{i}x_\mathrm{i,dq}$};
\node at (0.7,1.7) {$i_\mathrm{abc}$};

\draw (0.7,1.5) -- (1.1,1.5);
\node [black!50]at (-0.65,-1.45) {$v_\mathrm{c,dq}^\star$};
\draw [-latex](0.5,1.35) -- (0.9,1.35);
\draw [rounded corners = 2,black!50] (-1.1,-3.6) rectangle (0.7,-4.5);
\node [scale=0.9,black!50] at (-0.2,-3.75) {reactive power};
\node [scale=0.9,black!50] at (-0.2,-4) {PI control:};
\node [scale=0.9,black!50] at (-0.2,-4.3) {$k_\mathrm{p}^\mathrm{q}$, $k_\mathrm{i}^\mathrm{q}$};

\draw [rounded corners = 2,black!50] (-2.7,-3.6) rectangle (-1.3,-4.5);
\node [scale=0.9,black!50] at (-2,-3.75) {dc-voltage};
\node [scale=0.9,black!50] at (-2,-4) {PI control:};
\node [scale=0.9,black!50] at (-2,-4.3) {$k_\mathrm{p}^\mathrm{dc}$, $k_\mathrm{i}^\mathrm{dc}$};

\draw [-latex,black!50](-2,-3.6) -- (-2,-3.2); 
\draw [-latex,black!50](-0.2,-3.6) -- (-0.2,-3.2);
\node[black!50] at (-1.8,-3.4) {$i_\mathrm{d}^\star$};
\node [black!50]at (0,-3.4) {$i_\mathrm{q}^\star$};
\draw[-latex,black!50] (-2,-5) -- (-2,-4.5);
\node [scale=0.9] at (-1,1.96) {based VSC};
\draw [rounded corners = 2,black!50] (0.6,-0.2) rectangle (3.5,-1.4);
\node [scale=0.9,black!50] at (2.1,-0.4) {phase-locked-loop};
\draw [black!50] (-1.3,0.3) rectangle (-0.7,-0.2);
\draw [black!50](-1.3,-0.2) -- (-0.7,0.3); 
\node[scale=0.65,black!50] at (-1.1,0.17) {abc};
\node[scale=0.65,black!50] at (-0.9,-0.05) {dq};

\draw [black!50] (-4.3,-4.3) rectangle (-3.7,-4.8);
\draw [black!50](-4.3,-4.8) -- (-3.7,-4.3); 
\node[scale=0.65,black!50] at (-4.1,-4.43) {abc};
\node[scale=0.65,black!50] at (-3.9,-4.65) {dq};
\draw [-latex,black!50](-4.7,-4.55) -- (-4.3,-4.55); 
\draw[-latex,black!50](-3.7,-4.55) -- (-3.3,-4.55);

\draw [black!50] (-6.2,-4.3) rectangle (-5.6,-4.8);
\draw [black!50] (-6.2,-4.8) -- (-5.6,-4.3); 
\node[scale=0.65,black!50] at (-6,-4.43) {abc};
\node[scale=0.65,black!50] at (-5.8,-4.65) {dq};
\draw [-latex,black!50](-6.6,-4.55) -- (-6.2,-4.55); 
\draw[-latex,black!50](-5.6,-4.55) -- (-5.2,-4.55);
\node [black!50]at (-4.6,-4.3) {$i_\mathrm{abc}$};
\node [black!50] at (-6.5,-4.3) {$v_\mathrm{abc}$};
\node [black!50]at (-3.4,-4.3) {$i_\mathrm{dq}$};
\node [black!50] at (-5.3,-4.3) {$v_\mathrm{dq}$};

\draw [-latex,black!50](-5.9,-4) -- (-5.9,-4.3); 
\draw [-latex,black!50](-4,-4) -- (-4,-4.3); 
\node [black!50]at (-5.9,-3.8) {$\theta_\mathrm{pll}$};
\node [black!50] at (-4,-3.8) {$\theta_\mathrm{pll}$};

\node [black!50] at (1.3,-0.7) {$\dot{x}_\mathrm{pll}=v_\mathrm{q}$};
\node [black!50] at (2.1,-1.1) {$\dot{\theta}_\mathrm{pll}=k_\mathrm{p}^\mathrm{pll}v_\mathrm{q}+k_\mathrm{i}^\mathrm{pll}x_\mathrm{pll}$};
\draw[black!50,-latex] (1.1,-2.5) -- (0.5,-2.5);
\node [black!50] at (1.1,-2.7) {$v_\mathrm{dq}$, $i_\mathrm{dq}$};
\draw [black!50,-latex](2.1,-1.4) -- (2.1,-1.9);
\node [black!50] at (2.4,-1.7) {$\theta_\mathrm{pll}$};
\draw[black!50,-latex] (2.1,0.3) -- (2.1,-0.2);
\node [black!50] at (2.4,0.05) {$v_\mathrm{dq}$};
\draw [-latex,black!50](-0.2,0.05) -- (-0.7,0.05);
\node [black!50]at (0.1,0.05) {$\theta_\mathrm{pll}$};
\end{tikzpicture}
}
     \vspace{-7mm}
    \caption{One-line diagram of three-phase converter interface with matching control implementation. The converter model is in per unit, where $\mathcal{Z}_\mathrm{f}=L_\mathrm{f}\mathcal{J}_2+R_\mathrm{f}\mathcal{I}_2$ with $\mathcal{J}_2=[0\,\,\text{-}1;1\,\,0]$ and $\mathcal{I}_2=[1\,\,0;0\,\,1]$, and the associated parameters are provided in \cref{tab:converter_parameters}.}
    \label{fig:converter_model}
    \vspace{-1mm}
\end{figure}

\section{Transfer Function-Based Control of Converter-Interfaced Generation Systems} \label{sec:tf_based_control}\fontdimen2\font=0.6ex
The desired rational transfer function matrix $T_\mathrm{des}(s,\alpha)$ defines a grid-following frequency and voltage control behavior in the frequency domain which can be realized in a converter-based generation system, while ensuring grid-code and device-level requirements are satisfied. The proposed three-phase grid-side converter interface used for dynamic simulation is based on an average voltage source converter (VSC) model, and represents an aggregation of multiple commercial converter modules. As graphically illustrated in \cref{fig:converter_model}, we consider a state-of-the-art converter control scheme \cite{yazdani2010voltage}, into which we can easily incorporate the required transfer-function matching control. While \cref{fig:converter_model} shows only one exemplary converter control architecture, our method is compatible with any architecture that accepts active and reactive power references.

Similar to \cite{tayyebi2020frequency}, we assume the dc current $i_\mathrm{dc}$ to be supplied by a controllable dc current source, e.g., schematically representing the machine-side converter of a direct-drive wind power plant, a PV system, etc. In particular, we consider a generic coarse-grain model of the primary source technology and model its response time by a first-order delay with time constant $\tau_\mathrm{dc}$ \cite{tayyebi2020frequency}, e.g., representing the resource associated dynamics as well as communication and/or actuation delays. Moreover, we limit the primary source input by a saturation limit, which, e.g., corresponds to current limits of a machine-side converter or an energy storage system, or PV/wind power generation limits. In case of a wind or PV generation system, we assume they are operated under deloaded conditions with respect to their maximum power point, allowing them to put an active power reserve aside for participating in frequency regulation. While we stick to such an abstract representation of the primary source, one could of course consider more detailed models tailored to the specific application at hand. 

The ac-side control of the grid-side converter interface is implemented in a $\mathrm{dq}$-coordinate frame oriented via a phase-locked loop (PLL). The PLL is given by a state-of-the-art proportional-integral (PI) control to track the system frequency after the $RL$ filter, while keeping the converter synchronized with the grid voltage \cite{yazdani2010voltage}. Moreover, as indicated in \cref{fig:converter_model}, the inner current control is based on a conventional PI-control structure used to control the network current magnitudes. The associated current references are obtained from the outer control loops as described in the following.

Given the cascaded controller structure of the converter-interfaced generation system, we can directly incorporate the transfer function matching control in the outer control loops of the dc and ac side of the converter via simple PI controllers to track the desired dynamic response behavior for frequency and voltage regulation specified by $T_\mathrm{des}^\mathrm{fp}(s,\alpha)$ and $T_\mathrm{des}^\mathrm{vq}(s,\alpha)$, respectively\footnote{The practical implementation of the proposed transfer function matching control strategy has been experimentally validated in a classical power converter control setup in our recent work in \cite{andrejewski2023experimental}.}. Likewise, as indicated in light blue in \cref{fig:converter_model}, also classical (filtered) virtual inertia and droop specifications can be implemented in the same way. The stability of such a hierarchical converter control structure is guaranteed by time-scale separation, and can be analysed by standard methods, e.g., classical eigenvalue or Lyapunov analysis \cite{subotic2020lyapunov}. Finally, notice that alternatively, more robust and optimal matching controllers can be obtained by replacing the cascaded PI loops with a multivariable linear parameter-varying (LPV) $\mathcal{H}_\infty$ controller (see \cite{haberle2021control,haberle2022control}), which ensures device-level stability, and can additionally accomplish superior matching behavior especially in case of a parameter-varying $T_\mathrm{des}(s,\alpha)$ specification.
 
\section{Case Studies}\label{sec:case_studies}
\begin{figure}[t!]
 \centering
\begin{subfigure}{0.23\textwidth}
\centering
\hspace{-1mm}
    \scalebox{0.45}{\includegraphics[]{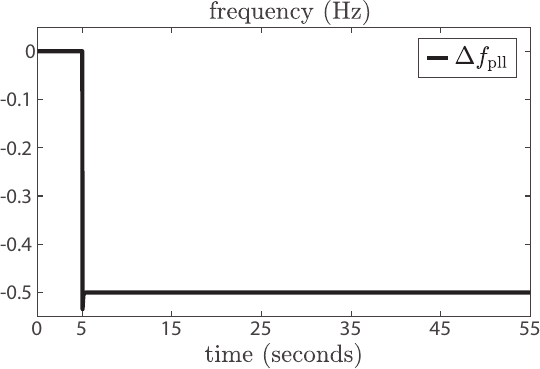}}
    \vspace{-0.5cm}
    \caption{Measured frequency step change at converter terminal.}
    \label{fig:f_pll}
\end{subfigure}
\hspace{1.8mm}
\begin{subfigure}{0.23\textwidth}
\centering
    \scalebox{0.45}{\includegraphics[]{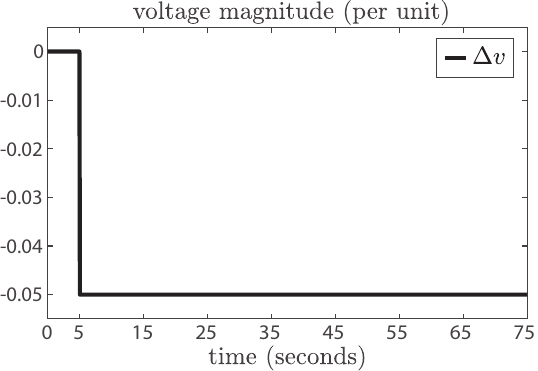}}
    \vspace{-0.5cm}
    \caption{Measured voltage magnitude step change at converter terminal.}
    \label{fig:v_pcc_mag}
\end{subfigure}
    \caption{Measured frequency and voltage magnitude step changes during the performed grid-code compliance tests.}
    \label{fig:input_signals}
    \vspace{-3mm}
\end{figure}

\fontdimen2\font=0.6ex
To demonstrate the performance of the proposed transfer-function based converter control, we consider a common setup for practical grid-code compliance tests \cite{oyj2021technical}, where the converter-interfaced reserve unit is connected to an infinite bus (\cref{fig:converter_model}). The parameters of the test setup used for the EMT simulations with MATLAB/Simulink are provided in \cref{tab:converter_parameters}. To evaluate the ability of the reserve unit in providing grid-following dynamic ancillary services for frequency and voltage regulation according to the active and reactive power time-domain capability curves specified in the grid-code, it is tested under some small-signal step change scenarios in frequency and voltage. Depending on the particular grid-code, the size of such step changes can be different. We thus consider a negative test frequency and voltage step change of $0.5$ Hz and $0.05$ p.u., respectively, where the measured input signals of the reserve unit are depicted in \cref{fig:input_signals}. However, our methods can be generalized to other step change amplitudes, and, beyond that, also be applied to actual power grid interconnections (see, e.g., \cite{haberle2024optimal} for an example).

\subsection{Tailored Transfer-Function Based Converter Control}\label{sec:tailored_tf_control_case_study}
\begin{figure}[t!]
    \centering
    \hspace{-5mm}
\begin{subfigure}{0.23\textwidth}
\centering
\hspace{-1mm}
    \scalebox{0.45}{\includegraphics[]{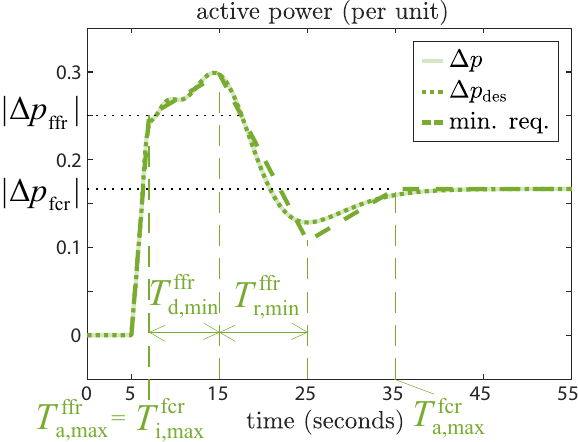}}
    \vspace{-0.5cm}
    \caption{Active power response.}
    \label{fig:min_grid_code_requirements_simulation_fp}
\end{subfigure}
\hspace{1.5mm}
\begin{subfigure}{0.23\textwidth}
\centering
    \scalebox{0.45}{\includegraphics[]{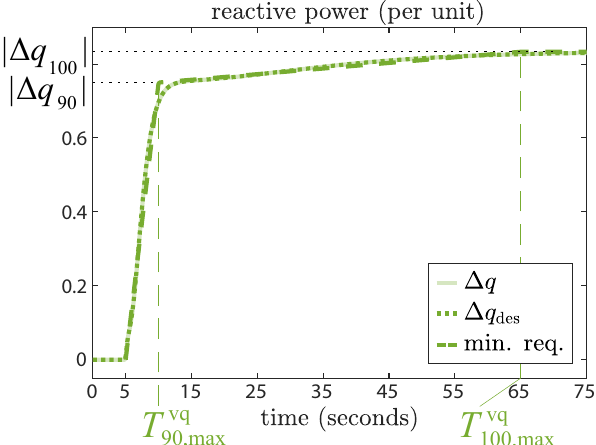}}
    \vspace{-0.5cm}
    \caption{Reactive power response.}
    \label{fig:min_grid_code_requirements_simulation_qv}
\end{subfigure}
    \caption{System response of the $T_\mathrm{des}(s,\alpha)$-based converter control during a negative frequency and voltage step ($\alpha$ satisfies \color{backgroundcolor2} min. grid-code requirements\color{black}).}
    \label{fig:min_grid_code_requirements_simulation}
    \vspace{-3mm}
\end{figure}

\begin{figure}[t!]
    \centering
    \hspace{-5mm}
\begin{subfigure}{0.23\textwidth}
\centering
\hspace{-1mm}
    \scalebox{0.451}{\includegraphics[]{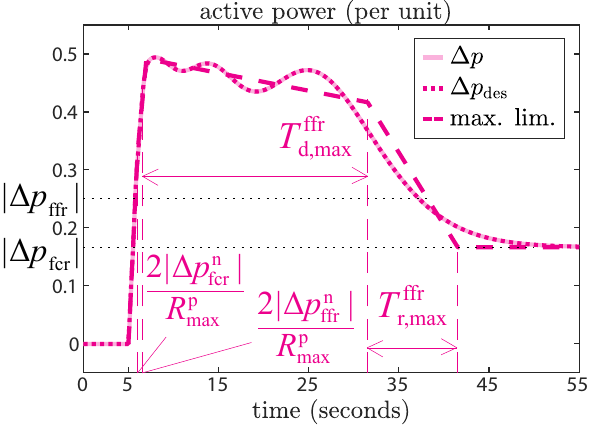}}
    \vspace{-0.5cm}
    \caption{Active power response.}
    \label{fig:max_device_level_limitations_simulation_fp}
\end{subfigure}
\hspace{1.5mm}
\begin{subfigure}{0.23\textwidth}
\centering
    \scalebox{0.45}{\includegraphics[]{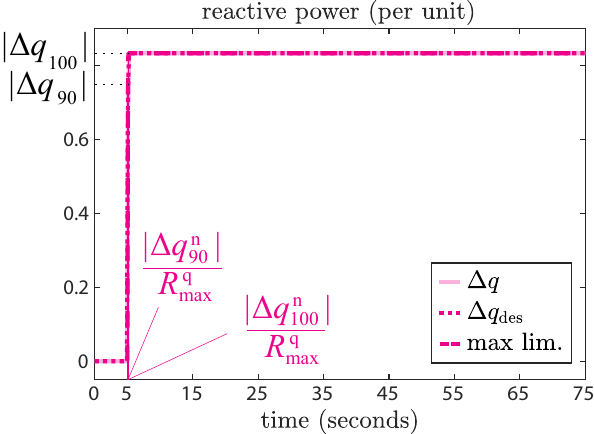}}
    \vspace{-0.5cm}
    \caption{Reactive power response.}
    \label{fig:max_device_level_limitations_simulation_qv}
\end{subfigure}
    \caption{System response of $T_\mathrm{des}(s,\alpha)$-based converter control during a negative frequency and voltage step ($\alpha$ exploits \color{magenta} max. device-level limitations\color{black}).}
     \vspace{-3mm}
    \label{fig:max_device_level_limitations_simulation}
\end{figure}

\fontdimen2\font=0.6ex
The test results of the two boundary scenarios presented in \cref{sec:parameter_selection} are illustrated in \cref{fig:min_grid_code_requirements_simulation,fig:max_device_level_limitations_simulation}. We can see how the measured active and reactive power deviations $\Delta p$ and $\Delta q$ accurately match their desired behaviors as specified by $T_\mathrm{des}^\mathrm{fp}(s,\alpha)$ and $T_\mathrm{des}^\mathrm{vq}(s,\alpha)$, respectively (dotted lines). Moreover, given the deliberate choice of $\alpha$, both grid-code and device-level requirements are reliably satisfied (dashed lines).

\subsection{Conventional Virtual Inertia \& Droop Control}\fontdimen2\font=0.6ex
Next, we perform the same grid-code compliance tests for a classical virtual inertia and droop control strategy \cite{poolla2019placement,ferreira2021control,stanojev2022improving} with a simple first-order filter. For the converter-interfaced generation system in \cref{fig:converter_model}, we can resort to the same PI-based matching control implementation as before, where, however, the desired dynamic response behavior is now described by the filtered virtual inertia and droop control (cf. light gray boxes in \cref{fig:converter_model}) accordingly as
\begin{align}\label{eq:inertia_droop_dynamics}
    \hspace{-2mm}\Delta p(s)=\tfrac{-Ms-1/D_\mathrm{p}}{\tau_\mathrm{f} s +1}\Delta f(s)\quad\,\Delta q(s)=\tfrac{-1/D_\mathrm{q}}{\tau_\mathrm{f} s +1}\Delta v(s),
\end{align}
where $M=4$, $D_\mathrm{p}=0.06$ are the virtual inertia and droop coefficients for the frequency regulation, $D_\mathrm{q}=0.06$ is a droop gain for the voltage regulation, and $\tau_\mathrm{f}\in\{0.1\text{s},\,2\mathrm{s}\}$ is the filter time constant. Notice that we use the same droop gains as encoded in $T_\mathrm{des}(s,\alpha)$ before (cf. \cref{tab:grid_code_parameters}), since they are typically directly given by the system operator.

\begin{figure}[t!]
    \centering
    \hspace{-6mm}
\begin{subfigure}{0.23\textwidth}
\centering
\hspace{-1mm}
    \scalebox{0.45}{\includegraphics[]{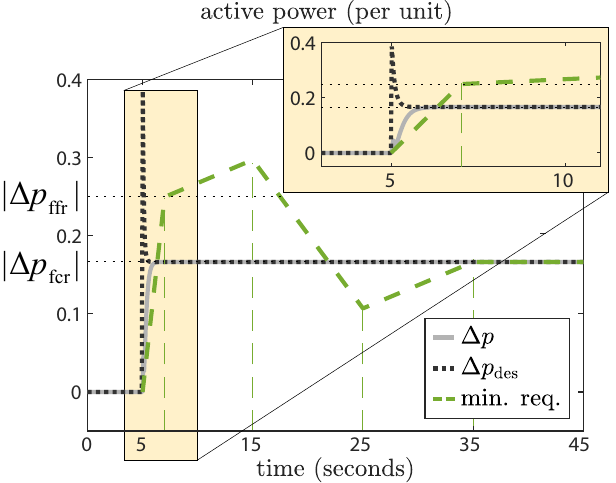}}
    \vspace{-0.5cm}
    \caption{Active power response.}
    \label{fig:inertia_droop_fast_fp}
\end{subfigure}
\hspace{1.8mm}
\begin{subfigure}{0.23\textwidth}
\centering
    \scalebox{0.45}{\includegraphics[]{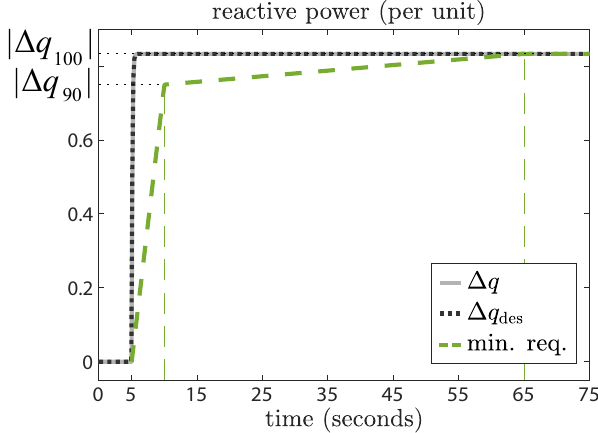}}
    \vspace{-0.5cm}
    \caption{Reactive power response.}
    \label{fig:inertia_droop_fast_qv}
\end{subfigure}
    \caption{System response of a filtered virtual inertia and droop control (filter time constant $\tau_\mathrm{f}=0.1$s) during a negative frequency and voltage step input.}
    \label{fig:inertia_droop_fast}
    \vspace{-3mm}
\end{figure}

\begin{figure}[t!]
    \centering
    \hspace{-6mm}
\begin{subfigure}{0.23\textwidth}
\centering
\hspace{-1mm}
    \scalebox{0.45}{\includegraphics[]{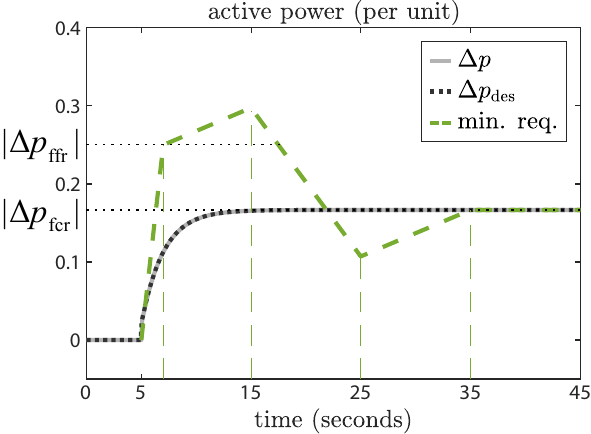}}
    \vspace{-0.5cm}
    \caption{Active power response.}
    \label{fig:inertia_droop_slow_fp}
\end{subfigure}
\hspace{1.8mm}
\begin{subfigure}{0.23\textwidth}
\centering
    \scalebox{0.45}{\includegraphics[]{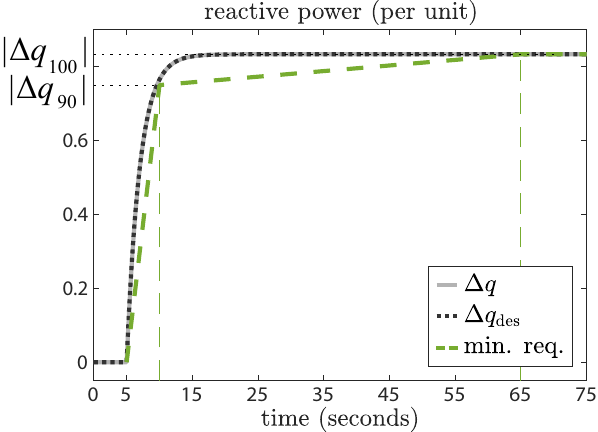}}
    \vspace{-0.5cm}
    \caption{Reactive power response.}
    \label{fig:inertia_droop_slow_qv}
\end{subfigure}
    \caption{System response of a filtered virtual inertia and droop control (filter time constant $\tau_\mathrm{f}=2$s) during a negative frequency and voltage step input.}
    \label{fig:inertia_droop_slow}
    \vspace{-4mm}
\end{figure}

By considering the desired active power response behavior in \cref{fig:inertia_droop_fast_fp,fig:inertia_droop_slow_fp} (dotted lines), we can observe how the dynamic ancillary services provision via filtered virtual inertia and droop control is not able to satisfy the \color{backgroundcolor2} minimum grid-code requirements (dashed lines)\color{black}. Namely, in contrast to $T_\mathrm{des}^\mathrm{fp}(s,\alpha)$ which allows for more structural flexibility to accomplish the grid-code requirements (see, e.g., the transfer function structure in \eqref{eq:Tdes_example_structure}), the demanded filtered virtual inertia and droop specifications in \eqref{eq:inertia_droop_dynamics} are limited in their fixed dynamic structure and thus often fail during grid-code compliance tests. In particular, this cannot be fixed by a proper tuning of the virtual inertia, droop or filter parameters. Instead, to overcome this problem, one would therefore require more sophisticated filters, which are aiming to replicate the structure of $T_\mathrm{des}^\mathrm{fp}(s,\alpha)$. On the other hand, due to the particular shape of the considered reactive power time-domain capability curve, the desired reactive power response of the filtered droop control (dotted lines) does satisfy the \color{backgroundcolor2} minimum grid-code requirements (dashed lines) \color{black} (\cref{fig:inertia_droop_fast_qv,fig:inertia_droop_slow_qv}). However, this is not guaranteed for other types of grid-code specifications.

\begin{figure}[t!]
   \centering
 \scalebox{0.5}{\includegraphics[]{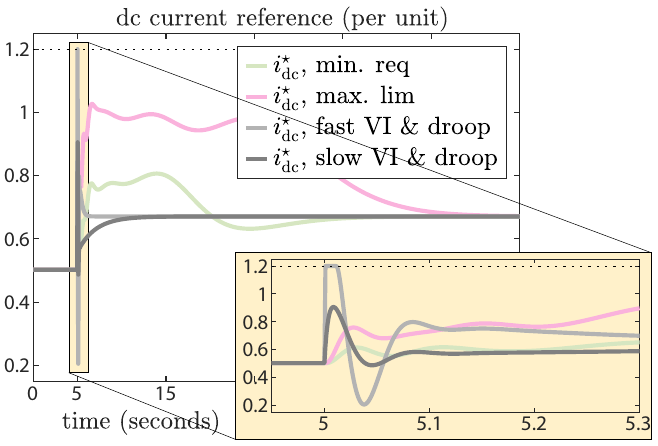}}
    \vspace{0.1cm}
    \caption{Dc-current reference signal for different types of converter control during the negative test frequency step change. Since the filtered virtual inertia (VI) and droop control does not account for the response-time limitations of the primary source, it can hit the dc-current reference saturation limit at 1.2 p.u.. The proposed transfer-function based converter control for the two boundary scenarios \color{backgroundcolor2} minimum grid-code requirements \color{black} (``min. req.'' curve) and \color{magenta} maximum device-level limitations \color{black} (``max. lim.'' curve) simulated in \cref{sec:tailored_tf_control_case_study}, in turn, can ensure such device-level limitations to be satisfied.}
    \label{fig:dc_current_ref}
    \vspace{-4mm}
\end{figure}

Finally, aside from violating grid-code requirements with the filtered virtual inertia and droop control, also device-level limitations can be hit if the filter time constant is not chosen appropriately. For example, as illustrated in \cref{fig:dc_current_ref}, the virtual inertia and droop control with the fast filter time constant is hitting the dc current reference saturation limit at 1.2 p.u. during a negative test frequency step change, as it does not account for the active power response-time limitation of the primary source. This results in a mismatch of the actually obtained (solid line) and the desired active power injection (dotted lines) in \cref{fig:inertia_droop_fast_fp}. Hence, the virtual inertia and droop control has to be deliberately slowed down, to ensure the device-level limitations are not violated (\cref{fig:inertia_droop_slow_fp}).

\section{Conclusion}\label{sec:conclusion}\fontdimen2\font=0.6ex
We have presented a systematic approach on how to translate piece-wise linear time-domain grid-code curves into desired rational parametric transfer functions in the frequency domain. The latter can be easily realized in standard converter control architectures to provide dynamic ancillary services, while ensuring grid-code and device-level requirements to be satisfied. Our numerical experiments verified the effectiveness of our proposed transfer function-based converter control, and especially revealed its superiority over classical droop and virtual inertia schemes, which are limited by their fixed controller structure and thus not able to satisfy a given grid-code requirement.

Finally, our proposed method is versatile to match any given piece-wise linear time-domain response curve, and thus also allows for more complex grid-code specifications in the future, e.g., also grid-forming grid-code requirements. On top of that, it might even inspire an immediate transfer-function based formulation of future grid-code requirements. 

\section*{Acknowledgements}
The authors wish to thank Simon Wenig from Mosaic Grid Solutions \cite{mosaic2023} for his fruitful comments and discussions.

\fontdimen2\font=0.4ex
\renewcommand{\baselinestretch}{0.94}
\bibliographystyle{IEEEtran}
\bibliography{IEEEabrv,mybibliography}

\end{document}